\input harvmac

\let\includefigures=\iftrue
\let\useblackboard=\iftrue
\newfam\black

\includefigures
\message{If you do not have epsf.tex (to include figures),}
\message{change the option at the top of the tex file.}
\input epsf
\def\figin{\epsfcheck\figin}\def\figins{\epsfcheck\figins}
\def\epsfcheck{\ifx\epsfbox\UnDeFiNeD
\message{(NO epsf.tex, FIGURES WILL BE IGNORED)}
\gdef\figin##1{\vskip2in}\gdef\figins##1{\hskip.5in}
\else\message{(FIGURES WILL BE INCLUDED)}%
\gdef\figin##1{##1}\gdef\figins##1{##1}\fi}
\def\DefWarn#1{}
\def\figinsert{\goodbreak\midinsert}
\def\ifig#1#2#3{\DefWarn#1\xdef#1{Fig.~\the\figno}
\writedef{#1\leftbracket Fig.\noexpand~\the\figno}%
\figinsert\figin{\centerline{#3}}\medskip\centerline{\vbox{
\baselineskip12pt\advance\hsize by -1truein
\noindent\footnotefont{\bf Fig.~\the\figno:} #2}}
\bigskip\endinsert\global\advance\figno by1}
\else
\def\ifig#1#2#3{\xdef#1{Fig.~\the\figno}
\writedef{#1\leftbracket Fig.\noexpand~\the\figno}%
\global\advance\figno by1}
\fi

\def\doublefig#1#2#3#4{\DefWarn#1\xdef#1{Fig.~\the\figno}
\writedef{#1\leftbracket Fig.\noexpand~\the\figno}%
\figinsert\figin{\centerline{#3\hskip1.0cm#4}}\medskip\centerline{\vbox{
\baselineskip12pt\advance\hsize by -1truein
\noindent\footnotefont{\bf Fig.~\the\figno:} #2}}
\bigskip\endinsert\global\advance\figno by1}

\useblackboard
\message{If you do not have msbm (blackboard bold) fonts,}
\message{change the option at the top of the tex file.}
\font\blackboard=msbm10 scaled \magstep1
\font\blackboards=msbm7
\font\blackboardss=msbm5
\textfont\black=\blackboard
\scriptfont\black=\blackboards
\scriptscriptfont\black=\blackboardss

\else

\fi
%
\def\subsubsec#1{\bigskip\noindent{\it{#1}} \bigskip}
\def\yboxit#1#2{\vbox{\hrule height #1 \hbox{\vrule width #1
\vbox{#2}\vrule width #1 }\hrule height #1 }}
\def\fillbox#1{\hbox to #1{\vbox to #1{\vfil}\hfil}}
\def\ybox{{\lower 1.3pt \yboxit{0.4pt}{\fillbox{8pt}}\hskip-0.2pt}}
%
%


\def\comments#1{}

\def\CN{{\cal N}}
\def\CO{{\cal O}}


\def\II{\relax{I\kern-.10em I}}

\def\IZ{\relax{\rm Z\kern-.34em Z}}
\def\IB{\relax{\rm I\kern-.18em B}}
\def\IC{{\relax\hbox{$\inbar\kern-.3em{\rm C}$}}}
\def\ID{\relax{\rm I\kern-.18em D}}
\def\IE{\relax{\rm I\kern-.18em E}}
\def\IF{\relax{\rm I\kern-.18em F}}
\def\IG{\relax\hbox{$\inbar\kern-.3em{\rm G}$}}
\def\IGa{\relax\hbox{${\rm I}\kern-.18em\Gamma$}}
\def\IH{\relax{\rm I\kern-.18em H}}
\def\II{\relax{\rm I\kern-.18em I}}
\def\IK{\relax{\rm I\kern-.18em K}}
\def\IP{\relax{\rm I\kern-.18em P}}

%

\def\inbar{\,\vrule height1.5ex width.4pt depth0pt}

\def\IR{\relax{\rm I\kern-.18em R}}

\def\simgt{\hskip0.05in\relax{
\raise3.0pt\hbox{ $>$
{\lower5.0pt\hbox{\kern-1.05em $\sim$}} }} \hskip0.05in}

%


%

\def\lp10{\ell_p^{10}}
\def\lp11{\ell_p^{11}}
\def\R11{R_{11}}

\def\frac#1#2{{#1 \over #2}}



\newdimen\tableauside\tableauside=1.0ex
\newdimen\tableaurule\tableaurule=0.4pt
\newdimen\tableaustep
\def\phantomhrule#1{\hbox{\vbox to0pt{\hrule height\tableaurule width#1\vss}}}
\def\phantomvrule#1{\vbox{\hbox to0pt{\vrule width\tableaurule height#1\hss}}}
\def\sqr{\vbox{%
  \phantomhrule\tableaustep
  \hbox{\phantomvrule\tableaustep\kern\tableaustep\phantomvrule\tableaustep}%
  \hbox{\vbox{\phantomhrule\tableauside}\kern-\tableaurule}}}
\def\squares#1{\hbox{\count0=#1\noindent\loop\sqr
  \advance\count0 by-1 \ifnum\count0>0\repeat}}
\def\tableau#1{\vcenter{\offinterlineskip
  \tableaustep=\tableauside\advance\tableaustep by-\tableaurule
  \kern\normallineskip\hbox
    {\kern\normallineskip\vbox
      {\gettableau#1 0 }%
     \kern\normallineskip\kern\tableaurule}%
  \kern\normallineskip\kern\tableaurule}}
\def\gettableau#1 {\ifnum#1=0\let\next=\null\else
  \squares{#1}\let\next=\gettableau\fi\next}

\tableauside=1.0ex
\tableaurule=0.4pt


 %
 %
 \def\eqnn#1{\xdef #1{(\secsym\the\meqno)}\writedef{#1\leftbracket#1}%
 \global\advance\meqno by1\wrlabeL#1}
 \def\eqna#1{\xdef #1##1{\hbox{$(\secsym\the\meqno##1)$}}
 \writedef{#1\numbersign1\leftbracket#1{\numbersign1}}%
 \global\advance\meqno by1\wrlabeL{#1$\{\}$}}
 \def\eqn#1#2{\xdef #1{(\secsym\the\meqno)}\writedef{#1\leftbracket#1}%
 \global\advance\meqno by1$$#2\eqno#1\eqlabeL#1$$}

\global\newcount\itemno \global\itemno=0
\def\itemized{\global\itemno=0}
\def\itemaut#1{\global\advance\itemno by1\noindent\item{\the\itemno.}#1}


\def\eg{{\it e.g.}}
\def\ie{{\it i.e.}}

\hyphenation{Di-men-sion-al}



\lref\othervacuarefs{
E.~J.~Martinec and G.~W.~Moore, ``On decay of K-theory,'' arXiv:hep-th/0212059;
G.~W.~Moore and A.~Parnachev,
``Localized tachyons and the quantum McKay correspondence,''
JHEP {\bf 0411}, 086 (2004)
[arXiv:hep-th/0403016];
I.~Melnikov and M.~R.~Plesser, ``The 
Coulomb Branch in Gauged Linear Sigma Models,''
arXiv:hep-th/0501238.}

\lref\AdamsSV{
A.~Adams, J.~Polchinski and E.~Silverstein,
``Don't panic! Closed string tachyons in ALE space-times,''
JHEP {\bf 0110}, 029 (2001)
[arXiv:hep-th/0108075].
}

\lref\RohmAQ{ R.~Rohm, ``Spontaneous Supersymmetry Breaking In Supersymmetric String Theories,'' Nucl.\ Phys.\ B
{\bf 237}, 553 (1984).
}

\lref\stromcon{
A.~Strominger,
``Massless black holes and conifolds in string theory,'' Nucl.\ Phys.\ B {\bf 451}, 96 (1995)
[arXiv:hep-th/9504090].
}

\lref\dealwisetal{
S.~P.~de Alwis, J.~Polchinski and R.~Schimmrigk, ``Heterotic Strings With Tree Level Cosmological Constant,''
Phys.\ Lett.\ B {\bf 218}, 449 (1989);
R.~C.~Myers, ``New Dimensions For Old Strings,'' Phys.\ Lett.\ B {\bf 199}, 371 (1987).
}

\lref\DaCunhaFM{
J.~Polchinski,
``A Two-Dimensional Model For Quantum Gravity,''
Nucl.\ Phys.\ B {\bf 324}, 123 (1989);
B.~C.~Da Cunha and E.~J.~Martinec, ``Closed string tachyon condensation
and worldsheet inflation,'' Phys.\ Rev.\ D {\bf 68}, 063502 (2003) [arXiv:hep-th/0303087];
E.~J.~Martinec,
``The annular report on non-critical string theory,''
arXiv:hep-th/0305148.
}

\lref\WittenGJ{ E.~Witten, ``Instability Of The Kaluza-Klein Vacuum,'' Nucl.\ Phys.\ B {\bf 195}, 481 (1982).
}

\lref\ShankarCM{ R.~Shankar and E.~Witten, ``The S Matrix Of The Supersymmetric Nonlinear Sigma Model,'' Phys.\
Rev.\ D {\bf 17}, 2134 (1978).
}

\lref\AhnGN{ C.~Ahn, D.~Bernard and A.~LeClair, ``Fractional Supersymmetries In Perturbed Coset Cfts And
Integrable Soliton Theory,'' Nucl.\ Phys.\ B {\bf 346}, 409 (1990).
}

\lref\FabingerJD{ M.~Fabinger and P.~Horava, ``Casimir effect between world-branes in heterotic M-theory,''
Nucl.\ Phys.\ B {\bf 580}, 243 (2000) [arXiv:hep-th/0002073].
}

\lref\openconfine{
P.~Yi, ``Membranes from five-branes and fundamental strings from Dp branes,'' Nucl.\ Phys.\ B {\bf 550}, 214
(1999) [arXiv:hep-th/9901159].
O.~Bergman, K.~Hori and P.~Yi, ``Confinement on the brane,'' Nucl.\ Phys.\ B {\bf 580}, 289 (2000)
[arXiv:hep-th/0002223].
A.~Sen, ``Supersymmetric world-volume action for non-BPS D-branes,'' JHEP {\bf 9910}, 008 (1999)
[arXiv:hep-th/9909062].
}

\lref\sen{
A.~Sen, ``Non-BPS states and branes in string theory,'' arXiv:hep-th/9904207.
}

\lref\tachtime{
T.~Okuda and S.~Sugimoto, ``Coupling of rolling tachyon to closed strings,'' Nucl.\ Phys.\ B {\bf 647}, 101
(2002) [arXiv:hep-th/0208196];
A.~Maloney, A.~Strominger and X.~Yin,
``S-brane thermodynamics,''
JHEP {\bf 0310}, 048 (2003)
[arXiv:hep-th/0302146];
 N.~Lambert, H.~Liu and J.~Maldacena, ``Closed strings from decaying D-branes,'' arXiv:hep-th/0303139;
D.~Gaiotto, N.~Itzhaki and L.~Rastelli,
``Closed strings as imaginary D-branes,''
Nucl.\ Phys.\ B {\bf 688}, 70 (2004)
[arXiv:hep-th/0304192].
}

\lref\otherRG{
J.~R.~David, M.~Gutperle, M.~Headrick and S.~Minwalla, ``Closed string tachyon condensation on twisted
circles,'' JHEP {\bf 0202}, 041 (2002) [arXiv:hep-th/0111212];
M.~Headrick, S.~Minwalla and T.~Takayanagi, ``Closed string tachyon condensation: An overview,'' Class.\ Quant.\
Grav.\  {\bf 21}, S1539 (2004) [arXiv:hep-th/0405064];
}

\lref\davetal{
D.~R.~Morrison, K.~Narayan and M.~R.~Plesser, ``Localized tachyons in C(3)/Z(N),'' JHEP {\bf 0408}, 047 (2004)
[arXiv:hep-th/0406039];
D.~R.~Morrison and K.~Narayan, ``On tachyons, gauged linear sigma models, and flip transitions,''
arXiv:hep-th/0412337.
}

\lref\chicago{
 J.~A.~Harvey, D.~Kutasov, E.~J.~Martinec and G.~W.~Moore, ``Localized tachyons and RG flows,''
arXiv:hep-th/0111154;
}

\lref\FQS{
D.~Friedan, Z.~Qiu and S.~H.~Shenker, ``Conformal Invariance, Unitarity And Two-Dimensional Critical
Exponents,'' Phys.\ Rev.\ Lett.\  {\bf 52}, 1575 (1984).
}

\lref\KMS{
D.~A.~Kastor, E.~J.~Martinec and S.~H.~Shenker, ``RG Flow In N=1 Discrete Series,'' Nucl.\ Phys.\ B {\bf 316},
590 (1989).
}

\lref\earlierSStach{
S.~Kachru, J.~Kumar and E.~Silverstein, ``Orientifolds, RG flows, and closed string tachyons,'' Class.\ Quant.\
Grav.\  {\bf 17}, 1139 (2000) [arXiv:hep-th/9907038].
}

\lref\ColemanBU{ S.~R.~Coleman, ``Quantum Sine-Gordon Equation As The Massive Thirring Model,'' Phys.\ Rev.\ D
{\bf 11}, 2088 (1975).
}

\lref\GreeneYB{ B.~R.~Greene, K.~Schalm and G.~Shiu, ``Dynamical topology change in M theory,'' J.\ Math.\
Phys.\  {\bf 42}, 3171 (2001) [arXiv:hep-th/0010207];
}

\lref\classtop{
P.~S.~Aspinwall, B.~R.~Greene and D.~R.~Morrison, ``Calabi-Yau moduli space, mirror manifolds and spacetime
topology  change in string theory,'' Nucl.\ Phys.\ B {\bf 416}, 414 (1994) [arXiv:hep-th/9309097];
E.~Witten, ``Phases of N = 2 theories in two dimensions,'' Nucl.\ Phys.\ B {\bf 403}, 159 (1993)
[arXiv:hep-th/9301042];
J.~Distler and S.~Kachru, ``(0,2) Landau-Ginzburg theory,'' Nucl.\ Phys.\ B {\bf 413}, 213 (1994)
[arXiv:hep-th/9309110].
J.~Distler and S.~Kachru, ``Duality of (0,2) string vacua,'' Nucl.\ Phys.\ B {\bf 442}, 64 (1995)
[arXiv:hep-th/9501111].
}

\lref\trapping{
L.~Kofman, A.~Linde, X.~Liu, A.~Maloney, L.~McAllister and E.~Silverstein, ``Beauty is attractive: Moduli
trapping at enhanced symmetry points,'' JHEP {\bf 0405}, 030 (2004) [arXiv:hep-th/0403001];
T.~Mohaupt and F.~Saueressig, ``Effective supergravity actions for conifold transitions,'' arXiv:hep-th/0410272.
L.~Jarv, T.~Mohaupt and F.~Saueressig, ``M-theory cosmologies from singular Calabi-Yau compactifications,'' JCAP
{\bf 0402}, 012 (2004) [arXiv:hep-th/0310174].
E.~Silverstein and D.~Tong, ``Scalar speed limits and cosmology: Acceleration from D-cceleration,'' Phys.\ Rev.\
D {\bf 70}, 103505 (2004) [arXiv:hep-th/0310221];
  A.~Lukas, E.~Palti and P.~M.~Saffin,
 ``Type IIB conifold transitions in cosmology,''
  Phys.\ Rev.\ D {\bf 71}, 066001 (2005)
  [arXiv:hep-th/0411033].
}

\lref\quantop{
B.~R.~Greene, D.~R.~Morrison and A.~Strominger, ``Black hole condensation and the unification of string vacua,''
Nucl.\ Phys.\ B {\bf 451}, 109 (1995) [arXiv:hep-th/9504145];
P.~Candelas and X.~C.~de la Ossa, ``Comments On Conifolds,'' Nucl.\ Phys.\ B {\bf 342}, 246 (1990);
S.~Kachru and E.~Silverstein, ``Chirality-changing phase transitions in 4d string vacua,'' Nucl.\ Phys.\ B {\bf
504}, 272 (1997) [arXiv:hep-th/9704185].
S.~Gukov, J.~Sparks and D.~Tong, ``Conifold transitions and five-brane condensation in M-theory on Spin(7)
Class.\ Quant.\ Grav.\  {\bf 20}, 665 (2003) [arXiv:hep-th/0207244].
}

\lref\MandelstamHB{
S.~Mandelstam,
``Soliton Operators For The Quantized Sine-Gordon Equation,''
Phys.\ Rev.\ D {\bf 11}, 3026 (1975).
}
\lref\ColemanBU{
S.~R.~Coleman,
``Quantum Sine-Gordon Equation As The Massive Thirring Model,''
Phys.\ Rev.\ D {\bf 11}, 2088 (1975).
}

\lref\KosterlitzXP{
J.~M.~Kosterlitz and D.~J.~Thouless,
``Ordering, Metastability And Phase Transitions In Two-Dimensional  Systems,''
J.\ Phys.\ C {\bf 6}, 1181 (1973).
}

\lref\KogutSN{
J.~B.~Kogut and L.~Susskind,
``Vacuum Polarization And The Absence Of Free Quarks In Four-Dimensions,''
Phys.\ Rev.\ D {\bf 9}, 3501 (1974).
}

\lref\PolyakovFU{
A.~M.~Polyakov,
``Quark Confinement And Topology Of Gauge Groups,''
Nucl.\ Phys.\ B {\bf 120}, 429 (1977).
}

\lref\AhnUQ{ C.~Ahn, ``Complete S Matrices Of Supersymmetric Sine-Gordon Theory And Perturbed Superconformal
Minimal Model,'' Nucl.\ Phys.\ B {\bf 354}, 57 (1991).
}

\lref\ZamolodchikovXM{ A.~B.~Zamolodchikov and A.~B.~Zamolodchikov, ``Factorized S-Matrices In Two Dimensions As
The Exact Solutions Of  Certain Relativistic Quantum Field Models,'' Annals Phys.\  {\bf 120}, 253 (1979).
}

\lref\PolchinskiFN{
J.~Polchinski,
``A Two-Dimensional Model For Quantum Gravity,''
Nucl.\ Phys.\ B {\bf 324}, 123 (1989).
}

\lref\Xiao{
S.~Hellerman and X.~Liu,
``Dynamical dimension change in supercritical string theory,''
arXiv:hep-th/0409071.
}

\lref\SaltmanJH{
A.~Saltman and E.~Silverstein,
``A new handle on de Sitter compactifications,''
arXiv:hep-th/0411271.
}

\lref\ScherkTA{
J.~Scherk and J.~H.~Schwarz,
``Spontaneous Breaking Of Supersymmetry Through Dimensional Reduction,''
Phys.\ Lett.\ B {\bf 82}, 60 (1979).
}

\lref\HollowoodEX{
T.~J.~Hollowood and E.~Mavrikis,
``The N = 1 supersymmetric bootstrap and Lie algebras,''
Nucl.\ Phys.\ B {\bf 484}, 631 (1997)
[arXiv:hep-th/9606116].
}

\lref\BajnokDK{
Z.~Bajnok, C.~Dunning, L.~Palla, G.~Takacs and F.~Wagner,
``SUSY sine-Gordon theory as a perturbed conformal field theory and  finite
size effects,''
Nucl.\ Phys.\ B {\bf 679}, 521 (2004)
[arXiv:hep-th/0309120].
}

\lref\FerraraJV{
S.~Ferrara, L.~Girardello and S.~Sciuto,
``An Infinite Set Of Conservation Laws Of The Supersymmetric Sine-Gordon
Theory,''
Phys.\ Lett.\ B {\bf 76}, 303 (1978).
}

\lref\WittenYC{
E.~Witten,
``Phases of N = 2 theories in two dimensions,''
Nucl.\ Phys.\ B {\bf 403}, 159 (1993)
[arXiv:hep-th/9301042].
}

\lref\VafaRA{
C.~Vafa,
``Mirror symmetry and closed string tachyon condensation,''
arXiv:hep-th/0111051.
}

\lref\MorrisonJA{
D.~R.~Morrison and K.~Narayan,
``On tachyons, gauged linear sigma models, and flip transitions,''
arXiv:hep-th/0412337.
}

\lref\BuscherQJ{
T.~H.~Buscher,
``Path Integral Derivation Of Quantum Duality In Nonlinear Sigma Models,''
Phys.\ Lett.\ B {\bf 201}, 466 (1988).
}

\lref\RocekPS{
M.~Rocek and E.~Verlinde,
``Duality, quotients, and currents,''
Nucl.\ Phys.\ B {\bf 373}, 630 (1992)
[arXiv:hep-th/9110053].
}

\lref\MorrisonYH{
D.~R.~Morrison and M.~R.~Plesser,
``Towards mirror symmetry as duality for two dimensional abelian gauge
theories,''
Nucl.\ Phys.\ Proc.\ Suppl.\  {\bf 46}, 177 (1996)
[arXiv:hep-th/9508107].
}

\lref\HoriKT{
K.~Hori and C.~Vafa,
``Mirror symmetry,''
arXiv:hep-th/0002222.
}

\lref\MorrisonYH{ D.~R.~Morrison and M.~R.~Plesser, ``Towards mirror symmetry as duality for two dimensional
abelian gauge theories,'' Nucl.\ Phys.\ Proc.\ Suppl.\  {\bf 46}, 177 (1996) [arXiv:hep-th/9508107].
}

\lref\AharonySX{
O.~Aharony, J.~Marsano, S.~Minwalla, K.~Papadodimas and M.~Van Raamsdonk,
``The Hagedorn/deconfinement phase transition in weakly coupled large N gauge
theories,''
arXiv:hep-th/0310285.
}

\lref\ZamolodchikovGT{
A.~B.~Zamolodchikov,
``'Irreversibility' Of The Flux Of The Renormalization Group In A 2-D Field
Theory,''
JETP Lett.\  {\bf 43}, 730 (1986)
[Pisma Zh.\ Eksp.\ Teor.\ Fiz.\  {\bf 43}, 565 (1986)].
}

\lref\BanksQS{
T.~Banks and E.~J.~Martinec,
``The Renormalization Group And String Field Theory,''
Nucl.\ Phys.\ B {\bf 294}, 733 (1987).
}

\lref\CostaNW{
M.~S.~Costa and M.~Gutperle,
``The Kaluza-Klein Melvin solution in M-theory,''
JHEP {\bf 0103}, 027 (2001)
[arXiv:hep-th/0012072].
}

\lref\GutperleMB{
M.~Gutperle and A.~Strominger,
``Fluxbranes in string theory,''
JHEP {\bf 0106}, 035 (2001)
[arXiv:hep-th/0104136].
}


\lref\GutperleBP{
M.~Gutperle,
``A note on perturbative and nonperturbative instabilities of twisted circles,''
Phys.\ Lett.\ B {\bf 545}, 379 (2002)
[arXiv:hep-th/0207131].
}

\lref\AspinwallRB{
P.~S.~Aspinwall, D.~R.~Morrison and M.~Gross,
``Stable singularities in string theory,''
Commun.\ Math.\ Phys.\  {\bf 178}, 115 (1996)
[arXiv:hep-th/9503208].
}

\lref\LindeSK{ A.~D.~Linde, ``Hard art of the universe creation (stochastic
approach to tunneling and baby
universe formation),'' Nucl.\ Phys.\ B {\bf 372}, 421 (1992)
[arXiv:hep-th/9110037].
}

\lref\ArkaniHamedSP{
N.~Arkani-Hamed, H.~Georgi and M.~D.~Schwartz,
``Effective field theory for massive gravitons and gravity in theory space,''
Annals Phys.\  {\bf 305}, 96 (2003)
[arXiv:hep-th/0210184].
}

\lref\VafaUE{
C.~Vafa,
``c Theorem And The Topology Of 2-D Qfts,''
Phys.\ Lett.\ B {\bf 212}, 28 (1988).
}

\lref\RobbinsHX{
D.~Robbins and S.~Sethi,
``A barren landscape,''
arXiv:hep-th/0405011.
}

\lref\PolchinskiHB{
J.~Polchinski,
``Cosmic superstrings revisited,''
arXiv:hep-th/0410082.
}

\lref\joequote{
``I had started to include a list,
but decided that it would be distracting
and inflammatory.  It was rather long.'' 
\PolchinskiHB.
}

\lref\friedan{
D.~Friedan,
``Nonlinear Models In $2 + \epsilon$ Dimensions,''
Phys.\ Rev.\ Lett.\  {\bf 45}, 1057 (1980).
}
\lref\Cappelli{
A.~Cappelli, D.~Friedan and J.~I.~Latorre,
``c-Theorem And Spectral Representation,''
Nucl.\ Phys.\ B {\bf 352}, 616 (1991).
}

\lref\ricciflow{
G.~Perelman, ``The entropy formula for the Ricci flow
and its geometric applications," arXiv:math.DG/0211159;
``Ricci flow with surgery on three-manifolds,"
arXiv:math.DG/0303109.
For a review of earlier work, see 
H.-D.~Cao, B.~Chow, ``Recent Developments on the Ricci Flow,"
arXiv:math.DG/9811123.}

\lref\HeadrickYU{
M.~Headrick,
JHEP {\bf 0403}, 025 (2004)
[arXiv:hep-th/0312213].
}

\lref\HeadrickPS{
M.~Headrick and J.~Raeymaekers,
arXiv:hep-th/0411148.
}
\lref\OkawaRH{
Y.~Okawa and B.~Zwiebach,
JHEP {\bf 0403}, 056 (2004)
[arXiv:hep-th/0403051].
}

\lref\timeprobes{
R.~Gregory and J.~A.~Harvey,
``Spacetime decay of cones at strong coupling,''
Class.\ Quant.\ Grav.\  {\bf 20}, L231 (2003)
[arXiv:hep-th/0306146]; 
M.~Headrick,
``Decay of C/Z(n): Exact supergravity solutions,''
JHEP {\bf 0403}, 025 (2004)
[arXiv:hep-th/0312213];
M.~Headrick and J.~Raeymaekers,
``The large N limit of C/Z(N) and supergravity,''
arXiv:hep-th/0411148;
Y.~Okawa and B.~Zwiebach,
``Twisted tachyon condensation in closed string field theory,''
JHEP {\bf 0403}, 056 (2004)
[arXiv:hep-th/0403051].
}

\lref\wormholerefs{ 
S.~R.~Coleman,
``Black Holes As Red Herrings: Topological Fluctuations And The Loss Of
Quantum Coherence,''
Nucl.\ Phys.\ B {\bf 307}, 867 (1988);
``Why There Is Nothing Rather Than Something: A Theory Of The Cosmological
Constant,''
Nucl.\ Phys.\ B {\bf 310}, 643 (1988);
S.~B.~Giddings and A.~Strominger, ``Loss Of Incoherence And Determination Of Coupling
Constants In Quantum Gravity,'' Nucl.\ Phys.\ B {\bf 307}, 854 (1988).
}

\lref\Barbon{
  J.~L.~F.~Barbon and E.~Rabinovici,
 ``Closed-string tachyons and the Hagedorn transition in AdS space,''
  JHEP {\bf 0203}, 057 (2002)
  [arXiv:hep-th/0112173];
 ``Remarks on black hole instabilities and closed string tachyons,''
  Found.\ Phys.\  {\bf 33}, 145 (2003)
  [arXiv:hep-th/0211212];
 ``Touring the Hagedorn ridge,''
  arXiv:hep-th/0407236;
 ``Topology change and unitarity in quantum black hole dynamics,''
  arXiv:hep-th/0503144.
}

\Title{\vbox{\baselineskip12pt\hbox{hep-th/0502021} \hbox{SU-ITP-05/06}\hbox{SLAC-PUB-11011}\hbox{HUTP-05/A0006}}}
{\vbox{
\centerline{Things Fall Apart:}\bigskip\centerline{Topology Change from 
Winding Tachyons}}}

\bigskip
\bigskip
\centerline{A. Adams$^1$, X. Liu$^2$, J. McGreevy$^2$, A. Saltman$^2$, E. Silverstein$^2$}
\bigskip

\centerline{$^1$ \it Jefferson Physical Laboratory, Harvard University, Cambridge, MA 02138}
\centerline{$^2$ \it SLAC and Department of Physics, Stanford University, Stanford, CA 94305-4060}
\bigskip
\bigskip
\noindent

We argue that closed string tachyons drive two spacetime topology
changing transitions -- loss of genus in a Riemann surface and
separation of a Riemann surface into two components. The tachyons
of interest are localized versions of Scherk-Schwarz winding
string tachyons arising on Riemann surfaces in regions of moduli
space where string-scale tubes develop. Spacetime and world-sheet
renormalization group analyses provide strong evidence that the
decay of these tachyons removes a portion of the spacetime,
splitting the tube into two pieces.  We address the fate of the
gauge fields and charges lost in the process, generalize it to
situations with weak flux backgrounds, and use this process to
study the type 0 tachyon, providing further evidence that its
decay drives the theory sub-critical. Finally, we discuss the
time-dependent dynamics of this topology-changing transition and
find that it can occur more efficiently than analogous transitions
on extended supersymmetric moduli spaces, which are limited by
moduli trapping.

\bigskip
\Date{January, 2004}

\newsec{Introduction and Setup}

One of the most basic aspects of 
quantum gravity is
the possibility of
dynamical change of spacetime topology.  In classical general
relativity such processes would be singular in the spacetime
metric, but in string theory the singularities may be smoothed out either
classically (via effects having to do with the extent of the
string) or quantum mechanically. Previous examples of the former
include \refs{\classtop,\davetal} and the latter \quantop.  In this note
we show that a natural conjecture for the endpoint of
condensation of a localized
tachyon
in string theory implies the
existence of two simple topology-changing processes. The
conjecture has substantial supporting evidence that we review and
develop in a similar spirit to \AdamsSV.

We compactify type II string theory down to eight dimensions on a
Riemann surface, which is topologically characterized by
the spin structure and Euler character
\eqn\eulerchar{\chi=2-2h,}
of each connected component, where the genus $h$ counts the number
of handles of a component. We will start with a single component.
Each handle adds energy density to the surface, as demonstrated by
the eight-dimensional Einstein frame potential energy of a
constant-curvature Riemann surface
\eqn\poten{U_{8E} \sim {1\over l_8^8}
\left( { g_s \over V_\Sigma^2 } \right)^{2/3} (2h-2),}
with 8-d Planck length $l_8$, string coupling $g_s$, and volume $V_\Sigma$, in string units.
The dependence on the dilaton and volume yields time-dependent
expansion and evolution toward weak coupling; we take the 
initial
volume large 
so that these effects are under control
in the low-energy effective theory.\foot{In \SaltmanJH, two
additional Riemann surface factors in the compactification
manifold and extra brane and flux ingredients were introduced to
perturbatively metastabilize this system in four dimensions,
leading to static solutions for the dilaton, complex structure,
and three volume moduli at large radius and weak coupling, away
from extreme limits of complex structure moduli space. Here we are
interested in a different regime of complex structure moduli
space, accessible from a simpler compactification on a single
Riemann surface down to eight dimensions, and we will necessarily
consider a time-dependent compactification as a result.} The
equations of motion for the metric in the low-energy theory yield
a mild FRW expansion, to which similar comments apply. Starting
from a more general metric of nonconstant curvature on the Riemann
surface, each volume element of negative curvature expands due to
the local energy \poten\ and each element of positive curvature
contracts. Further, the initial value of $g_s$ can be made small,
so that string interactions are negligible.

Because of the negative curvature of the compactification for $h
> 1$, this energy density \poten\ is positive and the system will tend to
reduce the genus if there is a dynamical mechanism by which it can
do so.\foot{In the superstring, a world-sheet Witten index valid
classically in the spacetime theory predicts further that there
will be additional discrete components to the target space, for
which we also find independent evidence.  Such effects have also
been seen before for example in \refs{\chicago,\othervacuarefs}.}
Factoring the Riemann surface into multiple components is also
energetically favored. If a Riemann surface of genus $h$ splits
into one of genus $0 < h_1 < h$ and another of genus $h-h_1$, the
Euler character on each of the resulting surfaces is smaller and
hence the potential energy \poten\ is smaller in each decoupled
sector than it used to be in the original connected space.  In all
cases, since we consider small string coupling the energy density
liberated from \poten\ is parameterically smaller than the Planck
energy density.
In this note, we will present evidence that both types of transitions are mediated by localized Scherk-Schwarz
winding tachyons.

\doublefig\before{ The two transition regions:  a) a thin handle;
b) a factorized surface. }{\epsfxsize2.0in\epsfbox{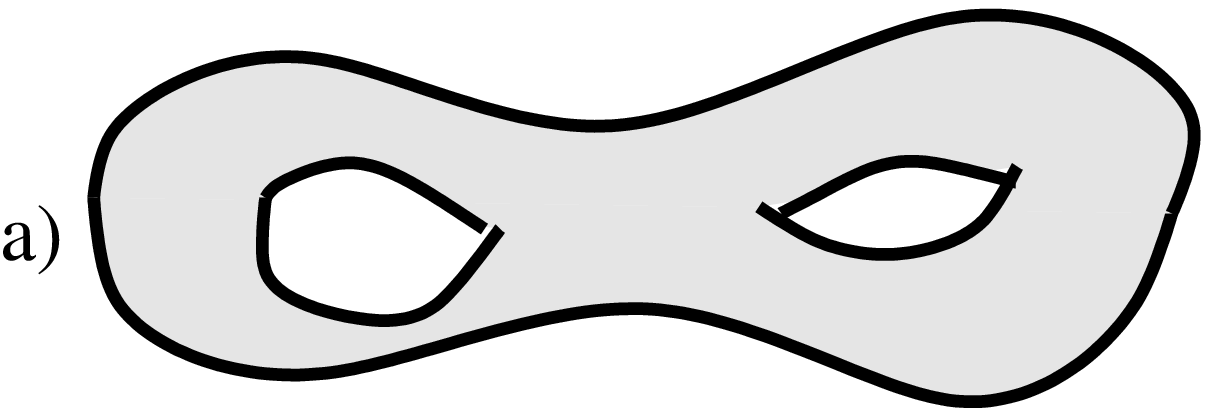}}
{\epsfxsize2.0in\epsfbox{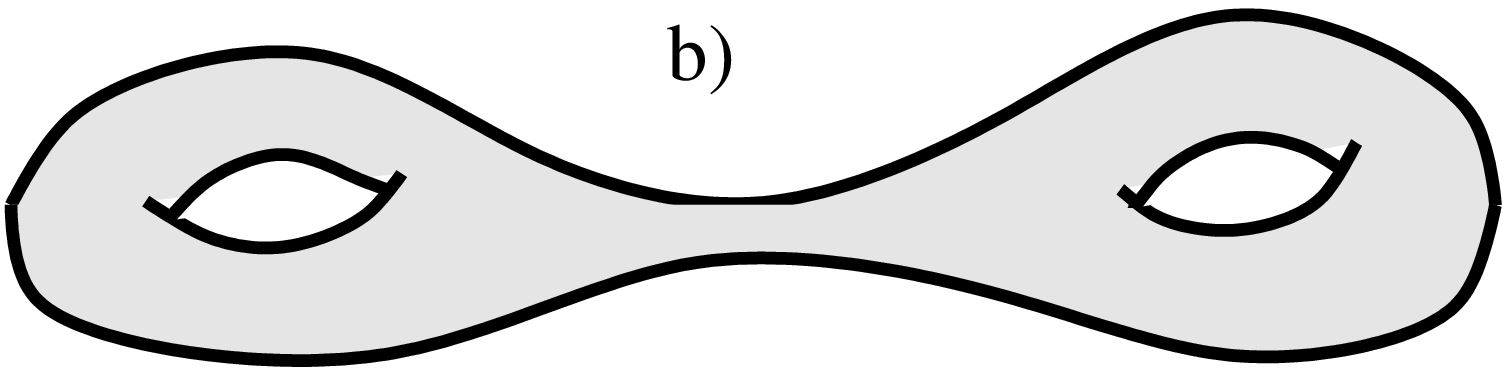}} We start from a
controlled regime in which curvatures are everywhere weak, but
where the Riemann surface degenerates in some local region to form
a long, thin (sub-string-scale) tube, with antiperiodic boundary
conditions for fermions around the circular direction. As depicted
in \before, this can happen when a handle degenerates or when the
surface nears a factorization limit. In the former case, in order
to ensure antiperiodic boundary conditions, we must choose the
spin structure appropriately.  In the latter case, the fermions
automatically have antiperiodic boundary conditions around the
thin cycle -- this can be seen by thinking of the Riemann surface
as a string world-sheet, in which case this statement is a
consequence of spacetime fermion number conservation.
We can consistently consider such regions while also maintaining small curvature (${\cal R} \alpha^\prime \ll
1$) everywhere: although a small length scale is developing on the thin tube, it is nearly flat.

The regions of complex structure moduli space containing small
tubes with the specified boundary conditions arise naturally from
perturbative dynamics.  Although classically the potential is flat
for the complex structure moduli (in the absence of stabilizing
fluxes \SaltmanJH), the 1-loop contribution to the potential
drives the radius of the tube to smaller values \RohmAQ.

The small circle introduces stringy physics -- classical geometric
notions break down. In particular, with the above specifications,
this tube is locally a Scherk-Schwarz (SS) circle \ScherkTA\ times
a line. Therefore, the theory develops tachyonic modes in its
winding string spectrum as the handle becomes thinner than the
string scale. For example, in ten-dimensional type II string
theory, the Scherk-Schwarz circle of radius $L l_s$ can be
obtained from a real line by an orbifold action $(-1)^F$ times a
translation by $L l_s$, where $F$ is the spacetime fermion number.
The world-sheet vacuum energy in the $n$th twisted sector is
$-1/2+n^2L^2$.
For sufficiently small $L$, this is negative, yielding tachyonic
modes in the spacetime spectrum.\foot{As $L\to 0$, these tachyonic
winding modes are T-dual to momentum modes of the Type 0 bulk
tachyon, whose condensation we will also examine below using
similar methods.}

\doublefig\after{
The centre cannot hold: a) a thin handle capped off; b) a
factorized surface capped off.
}{\epsfxsize2.0in\epsfbox{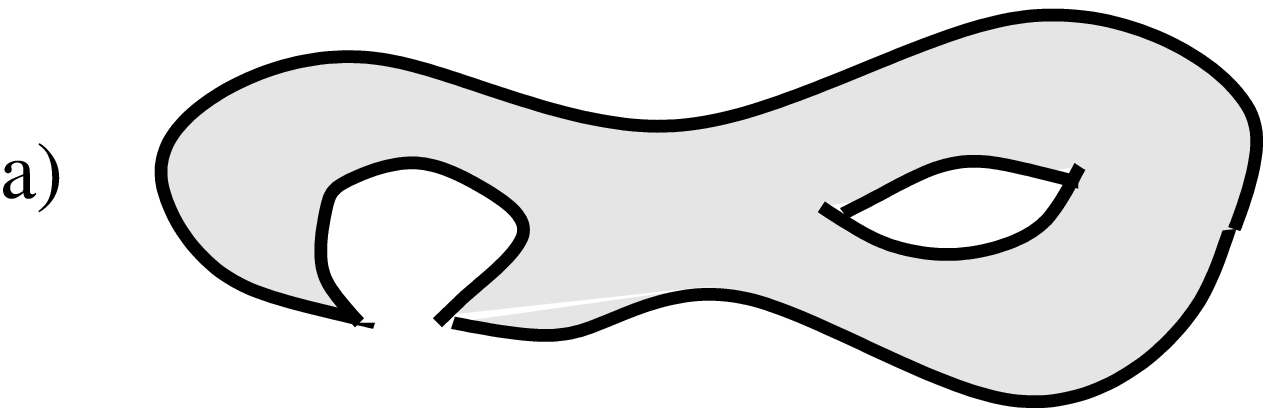}}
{\epsfxsize2.0in\epsfbox{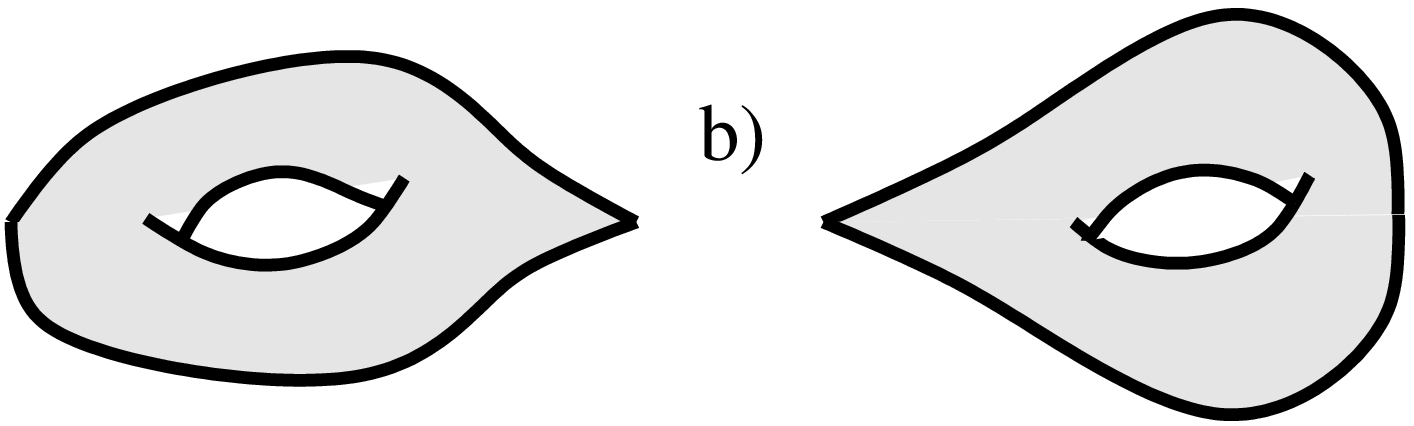}} Tachyon condensation
generically reduces the number of degrees of freedom, and in the
case of the Scherk-Schwarz tachyon we will show that this effect
follows from standard results concerning the mass gap of a
corresponding supersymmetric sine-Gordon theory in two dimensions.
Including both dimensions of the tube, we argue in a consistent
approximation scheme that condensing this tachyon causes the
system to lose the region where the circle is smaller than string
scale, breaking the tube there and capping off the remaining
regions (see \after). Our specification of small curvatures
everywhere guarantees that the time evolution induced by the
spacetime curvature term \poten\ is parameterically slower than
that of the tachyon decay process (at least initially).

That bulk Scherk-Schwarz tachyon condensation yields a subcritical
theory has been conjectured before in general terms
\refs{\dealwisetal,\earlierSStach, \FabingerJD}.  Here we analyze
this idea in detail in the more controlled setting described
above, in which the tachyon is localized, so that we can apply the
result to obtain the topology-changing processes just described.
The suggestion that Hagedorn tachyons
have an endpoint similar to the one described here 
has been made in \Barbon.
An interesting comment about winding tachyons 
in nonlinear sigma models was made in \friedan.

\subsec{Setup and Plan}

Classically condensing the tachyon in real time requires adding to
the world-sheet action the on-shell marginal tachyon vertex
operator and solving the resulting path integral in the deep IR.
In type II string theory, the world-sheet has (1,1) local
supersymmetry.  The zero ghost picture tachyon vertex operator is
of the form
\eqn\Top{\int d^2 \sigma d\theta^+ d\theta^- ~T(X)}
where we work in ${\cal N}=1$ superspace, with spacetime embedding coordinates $x$ extended to $\CN=1$ scalar
multiplets $X=x+\theta^+\psi^-+\theta^-\psi^+ + \theta^+\theta^- F$.

We denote the direction around the Scherk-Schwarz circle $\theta$,
its T-dual $\tilde\theta$, and the superspace coordinate
corresponding to the latter $\tilde\Theta$. We also parameterize
the direction lengthwise along the tube by $r$, with corresponding
superspace coordinate $R$.  Finally we denote the target-space
time by $t$, extended to superspace coordinate $X^0$.  The tachyon
vertex operator for our system is
\eqn\Tsupvert{T(X)=e^{\kappa X^0}\hat T(R)\cos[w\tilde\Theta],}
where $w=nL/l_s$ with $n$ the winding number and $L$ the radius of the tube.  The second reference in
\DaCunhaFM\ studied aspects of this system for other applications, and we will use some of its observations in
our analysis.

As we will discuss in more detail below, if we add a tachyon vertex operator with  mild $r$ and $t$ dependence,
\eqn\regime{\kappa^2 < k_r^2 \ll w^2}
the action generated by \Tsupvert\ is the supersymmetric sine-Gordon model (SSG) for $\tilde\Theta$, corrected
by subleading pieces depending on $\del_{X^0}T$ and $\del_r T$.

Classically, the world-sheet theory with the interaction terms
\Tsupvert\ has a mass gap for $\tilde\Theta$ sector in the region
of the tachyon condensate. The physics of the tachyon condensation
process is governed by the quantum theory in the IR limit. In the
regime \regime\ we will see that we can treat the IR quantum
dynamics of the SSG theory as the leading physics of the tachyon
condensation in the $\tilde\Theta$ sector.  As we will review,
there is strong evidence for a mass gap in this system.

In addition to its role in the full time-dependent problem induced
by the condensation of $T$ \Tsupvert, the renormalization group
(RG) flow in the world-sheet matter sector (which is induced by
the addition of the $\kappa=0$ vertex operator to the world-sheet
Lagrangian) may well reflect part of the off-shell configuration
space of string theory, since string field theory is built on
off-shell string states and vertices which sample non-conformal
regimes in the world-sheet matter theory. This has been argued in,
for example
\refs{\BanksQS,\VafaUE,\dealwisetal,\AdamsSV,\VafaRA,\otherRG,\chicago};
in all known cases the RG results agree with the pattern found by
probes of the time-dependent process
\refs{\sen,\AdamsSV,\timeprobes}.\foot{One can consider limits in which the
second order equations governing the time-dependent evolution of
the tachyon $T$ become first order, via a large coupling of the
form $\dot\Phi\dot T$ where $\Phi$ is some combination of dilaton
and volume moduli \otherRG. However, this approximation can break
down at later stages in the process.}

We thus start in \S2\ by reviewing the RG flow of the sine-Gordon
theory (supersymmetric and otherwise), as well as the RG flow of
this theory weakly coupled to $R$, in the matter theory on the
world-sheet ignoring the effects of $X^0$.  We see that the RG
flow removes the space in the region of the tachyon condensate. In
the appendix, we present a linear sigma model which exhibits the
flow for both dimensions at once; the model has a relevant
operator which changes the vacuum manifold from a connected
hyperboloid to a two-sheeted disconnected one. We also apply this
result to the problem of the condensation of the Type 0 tachyon.
We then apply the RG result to the classical time-dependent
problem \Tsupvert\ in \S3, exhibiting a barrier to penetration of
the world-sheet into the region of tachyon condensate. We also
comment on the efficiency of our topology change process
(including effects of nonzero string coupling), comparing and
contrasting to earlier tachyon and topology change analyses.

In \S4\ we discuss the fate of the winding charge when a homology
cycle is destroyed by tachyon condensation. We find that the gauge
field under which the condensing tachyon is charged is Higgsed and
that the dual gauge field appears to confine. In addition, we
comment on the behavior of D-brane charges under the tachyon
condensation, and extend our discussion to Riemann surfaces with
weak flux backgrounds.  In \S5\ we discuss some of the potential
implications of our results and possible directions for further
research.  In appendix A we 
describe a linear sigma model 
which captures the RG.

While this paper was in preparation we 
noticed that our results
had been anticipated by the 
prescient early work of Yeats.
In appendix B, we reproduce his argument,
with a translation to more
modern notation.
Note that his analysis applies
under more general circumstances.

\newsec{Renormalization Group Structure}

In order to study the RG problem in the matter sector, we suppress the $X^0$ dependence in the vertex operator
\Tsupvert. The resulting relevant operator in the region of the tube is (in the zero ghost picture) of the form
\eqn\Top{\int d^2 \sigma d^2\theta ~T(X)}
where we work in ${\cal N}=1$ superspace, with world-sheet fields
$x$ extended to $\CN=1$ scalar multiplets $X$.  We denote the
direction around the Scherk-Schwarz circle by $\theta$ (not to be
confused with the world-sheet superspace coordinates
$\theta^\pm$!), its T-dual by $\tilde\theta$, and the superfield
corresponding to the latter by $\tilde\Theta$. We also denote the
direction lengthwise along the tube by $r$, with corresponding
superfield $R$. The tachyon vertex operator at zero momentum is
\eqn\TsupvertII{T(X)=\hat T(R)\cos[w\tilde\Theta].}
\foot{Note that the winding tachyon is actually always complex,
because we could add the winding and antiwinding modes with
different phases
$$ T e^{ i w \tilde \theta} + \bar T e^{- i w \tilde \theta} $$
while preserving reality of the action. This phase in $T = e^{i
\alpha} |T|$ changes the cosine in \TsupvertII\ to $ \cos \left[ w
\tilde \Theta + \alpha\right] $. }Before proceeding, it is worth
commenting further on the supersymmetry structure of the tachyon
vertex.  In the case of \AdamsSV, the lowest-dimensional
zero-momentum tachyon vertex operators were chiral primaries, and
thus preserved ${\cal N}=2$ supersymmetry, leading to useful
${\cal N}=2$ linear sigma model treatments of the problem
\refs{\VafaRA,\otherRG}.  In our case, the matter sector vertex
operator \TsupvertII\ is not chiral primary except in the limit
$L=0$.  In the appendix, we describe a linear sigma model which
clarifies some physical features, but here we focus mostly
directly on the physical theory on the world-sheet.

In the $\theta$ direction, this operator is the interaction Lagrangian of the ${\cal N}=1$ supersymmetric
sine-Gordon theory, while the unperturbed action contains the kinetic term for $\theta$.  In components, the
classical action is
\eqn\SSGaction{\int d^2\sigma\left( g_{\tilde\theta
\tilde\theta}(r) \del_a\tilde\theta\del^a\tilde\theta +
g_{rr}(r)\del_a r\del^a r + \hat T(r)
\sin(\tilde\theta)\psi_{\tilde \theta}\tilde\psi_{\tilde\theta} +
\hat T^2(r) \cos^2\tilde\theta\right)}
where $\hat T(r)$ 
and $g_{ab}(r)$ vary slowly with $r$ near the point $r_0$ in the middle of the tube where
$g_{\theta\theta}$ reaches its minimum value.

Because the system varies only mildly with changes of $r$, let us
start by approximating $g$ and $T$ as independent of $r$, and
study the flow of the ${\cal N}=1$ supersymmetric sine-Gordon
theory in the $\tilde\Theta$ direction. Classically, all modes
(elementary and solitonic) have masses from the tachyon vertex. In
the quantum theory, there is compelling evidence that a mass gap
persists. The model is integrable \FerraraJV, and the well-checked
proposal for its exact factorized S-matrix
\refs{\ShankarCM,\AhnUQ,\HollowoodEX,\BajnokDK}
contains no massless singularities. Physically, this mass gap
results from a generalization of the Kosterlitz-Thouless
transition, in which vortex condensation destroys long range
order. To clarify the important aspects of this phenomenon, we
give a brief review of how it works in the bosonic case. This
analysis can be directly applied to winding tachyons in the
bosonic string if one is willing to tune to zero the omnipresent
bulk tachyon.

\subsec{A brief review of vortex-induced confinement}

The XY model provides an excellent demonstration
of the aphorism
that magnetic higgsing is electric confinement. Models in this
universality class are two-dimensional and have a $U(1)$ symmetry,
and therefore include a $U(1)$ Goldstone boson $\theta(z)$, of
some radius $L$, in their low energy description. We will describe
this phenomenon from the point of view of the line of fixed points
parametrized by the radius of the circle in units of the self-dual
radius\foot{In conventional string theory normalization $ S = {1
\over 4 \pi l_s^2 } \int d^2 \sigma \del \theta \bar \del \theta
$.}, $L_c = \sqrt 2 l_s$ (in the realization as a statistical
mechanics problem, this parameter can be traded for the
temperature). Single-valued winding and momentum modes of a boson
at radius $L$ ($\theta \simeq \theta + 2 \pi L$ ) are created by
the operators \eqn\windingmomentum{ \CO_{n,m} \equiv \exp{ i\left(
{ n\over L}  \theta + mL  \tilde \theta \right)} , ~~ m,n \in \IZ.
} This operator has conformal dimension $
\Delta_{n,m} = \left({n \over L}\right)^2 +  ( m L ) ^2 .
$
Therefore, when $L < L_c $,
there is a {\it relevant} winding operator $\CO_{0, \pm 1}$, with
(chiral) conformal dimension
$ \Delta_{0,\pm 1} = L^2  < 1 .$

The XY model provides an example of a system with `topological order' \KosterlitzXP: in the low-temperature
phase, although there is no local order parameter, correlators fall off as a power law with distance. Above a
critical temperature, 
this order is destroyed by the condensation of {\it disorder} operators, $\CO_{0,1}, \CO_{0,
-1}$ -- in the presence of the vertex operator for a winding mode, the $\theta$ field is constrained to have a
discontinuity. A gas of such insertions destroys the long-range correlations.

The quickest way to see that the condensation of vortices induces
a nonzero correlation length is to use the fact that this model
can be fermionized \refs{\MandelstamHB, \ColemanBU}. The operator
$\del \theta \bar \del\theta$ which changes the radius of $\theta$
fermionizes to a Thirring four-fermion operator; at a radius other
than the aptly named free-fermion radius, $L = L_{{\rm ff}}=\sqrt
2 L_c$, the fermions interact via a critical four-fermi term, but
are massless. In this Thirring description, the neutral winding
deformation $T \cos \tilde \theta$ is simply the fermion mass
operator.  Starting close to the limit of a free massive fermion,
this makes it clear that the IR limit that one reaches
by perturbing by the corresponding deformation
by $T$ is a massive theory (a result which also
obtains for any value of the mass). 

It is worth noting that in the case of $\IC^q/\IZ_N$ orbifolds 
\AdamsSV, the twisted tachyons at the orbifold singularities are
winding modes when pulled away from the tip.  For large $N$
this allows us to model the twisted tachyon vertex operator to
a good approximation in this region as a SSG winding operator.
Our analysis here of the corresponding mass gap thus provides
a useful corroboration of the decay seen in \AdamsSV\ via 
a more explicit analysis in the physical worldsheet theory.

\subsec{Return to the superstring}

The claim that the supersymmetric sine-Gordon
model flows to a theory with a mass gap
 can also be
supported
via a process of elimination as follows.  The operator \Top\ is
relevant at small enough radius, and hence its condensation
reduces the central charge in the $\theta$ direction from 3/2 to
something smaller. The representation content of all unitary
${\cal N}=1$ SCFTs with $c<3/2$ have been classified \FQS, and
all explicit examples 
include relevant operators preserving supersymmetry.  As long as some relevant operators survive under the
GSO projection applicable in our theory (and they do survive under the basic type II and type 0 GSO projections
manifest in the Landau-Ginzburg description \KMS), the system has no possible nontrivial infrared stable
endpoint.

\subsubsec{Dependence on $r$}

Now let us reinstate the $r$-dependence in our problem.  This can
be done by expanding $g_{ab}$ and $T$ in a Fourier series in the
region of the tube (equivalently in a derivative expansion with
respect to $r$).  The zero momentum term yields the flow just
described in the theta direction.  This indeed dominates as we
flow toward the infrared, since including the factor
$:e^{ik_rr}:$, the higher order terms involve higher dimension
operators than the leading term in the expansion in $k_r$.

Thus the $\Theta$ direction is removed by the flow.  Once this has
occurred, the remaining theory is subcritical and contains other
bulk tachyons (relevant operators on the world-sheet).  These
reduce the central charge further--in particular, in the $r$
direction the tachyonic modes appearing for sufficiently small
$k_r$ condense, again producing a theory of central charge $c<3/2$
which must flow to a trivial fixed point.  These deformations also
involve supersymmetric sine-Gordon interactions locally.

As we move away from the minimum $r_0$, we come to a value of $r$
at which the tachyon gradient ${\del T \over \del r}$ becomes
non-negligible. In this region there the capping-off process we
have inferred from the mass gap must occur. In the appendix, we
present a linear sigma model involving both the $R$ and $\Theta$
directions which exhibits a flow from a connected tube to two
disconnected copies of the complex plane. This confirms the result
argued here directly in the physical theory.

Once this process has removed the $R,\Theta$ degrees of freedom in the middle of the tube, the tube has split as
indicated in \before\ and \after.
In one case a handle has been lost (changing the Euler character of the Riemann
surface), and in the other the space has split into two disconnected parts (changing the most basic topological
invariant: the number of connected components).

Because of the invariance of the Witten index of the world-sheet
theory in this process, the change in Euler character in the
Riemann surface must be accompanied by the generation of other
isolated vacua.  This is evident from linear sigma model
descriptions of the process, as we discuss in the
appendix \AspinwallRB, and also emerges in our analysis of the
time-dependent effects of the tachyon condensation in the
next section.

\subsubsec{Remnants}

Our arguments do not rule out the possibility that there is some
nontrivial string scale fixed point, crucially involving both the
$r$ and $\theta$ degrees of freedom, at which the
handle-destruction process can abort. The theory would have to
take some trajectory other than the one above, which led to the
trivial IR fixed point in the region of the tachyon condensate. As
we just discussed, the flow of the sine-Gordon model itself is
well understood, so this possibility can be ruled out if the
$\theta$ direction does not mix with others.
However, it is logically possible that there are other
trajectories leading to some nontrivial fixed point for the
$r$-$\theta$ theory. It is only in the case that this fixed point
is IR stable (including projection by the appropriate GSO action)
that this could arise as a stable endpoint of the process for
appropriate initial conditions. An alternative trajectory and
endpoint would be fascinating in its own right, but in the absence
of an example of such a fixed point, we will focus here on the
generic flow to the trivial theory.

\subsec{c-curity Check and the Type 0 Tachyon}

The nonlinear sigma model (NLSM) on a {\it compact} Riemann
surface satisfies the hypotheses of Zamolodchikov's $c$-theorem
\refs{\ZamolodchikovGT, \Cappelli}, 
if we define the theory with a cutoff. On the
other hand, we claim the dimension of the target space remains two
after perturbation by the winding tachyon, a relevant operator.
Does this imply that our conjectured RG flow from the NLSM on a
compact Riemann surface with genus $h$ to one with genus $h-1$
(plus extra states elsewhere) violates the c-theorem?  The answer
is negative because the $c$-function is not simply the
dimensionality of the target space in the presence of curvature.

We distinguish two cases, $h > 1$ (the case considered above) and
$h=1$ (closely related to Type 0), and discuss them in turn.

\subsubsec{$h>1$}

While the flow induced by tachyon condensation from genus $h$ to $h-1$ leaves the dimension of spacetime fixed
at 2, it does not leave the $c$-function fixed: since the Riemann surface is not flat, the NLSM pertaining to
the matter sector alone is not conformal. The effective $c$ for a Riemann surface target space, ignoring the
$X^0$ direction, is greater than 2 and a monotonically decreasing function of scale, unless $h-1=1$. In
particular, it is never 2 (or $\hat 2$) when the volume is finite for a higher genus Riemann surface.

This is analogous to the familiar case of positively curved target spaces, for which the effective $c$ is less
than the dimensionality of the target space.  For example, in the NS5-brane solution there is an $S^3$ component
stabilized by fluxes whose $\hat c$ is less than 3; there the spacelike linear dilaton makes up this deficit. In
our case, time-dependence of the dilaton plays a similar role\foot{
Unlike our case, however, 
the WZW model is actually conformal, 
so its c-function is just the central charge.}

The $c$-function for the NLSM on a Riemann surface 
of genus $h$
and constant curvature $R \sim (2-2h)/V$, with volume
$V = v V_0$
(where $V_0$ is the volume of
some fiducial metric, and $v$ is a dimensionless coupling)
in perturbation theory around the free fixed point
at $v \to \infty$ is
\eqn\cfunction{
c = c_0 + b \alpha' R  + \CO(\alpha'R)^2,
}
where $c_0$ is the central charge at the fixed point,
and $b$ is a positive constant.
This statement applies both for the bosonic and for the 
supersymmetric NLSM,
though hats must be sprinkled appropriately on the RHS in the latter case.
This follows directly from equation (12)
of \ZamolodchikovGT, 
which gives the c-function in a perturbation expansion
about a weakly-coupled fixed point.

There are several possible endstates of such tachyonic decays.  After decay from genus $h$ to $h-1$, fluxes and
branes wrapping surviving cycles might stabilize the moduli of an $h>1$ endstate \SaltmanJH\ in regions of
complex structure moduli space without small Scherk-Schwarz handles, terminating the perturbative decay (though
non-perturbative effects may destabilize these moduli, sowing the seeds of further decay). In the absence of
such stabilizing effects, the end of the process depends on the spin structure on all the cycles; if a torus
remains with periodic boundary conditions, the RG flow can end with a flat metric on the resulting $h=1$
surface, yielding $\hat c=2$. It can also end at infinite volume ($\hat c =2$) via the overall flow toward large
volume at higher genus, if the unstable handles remain large and the system stays far from factorization limits.

However, as we discuss next, if there are remaining antiperiodic
boundary conditions for $h=1$, the system can decay further to a
genus 0 surface (a 2-sphere), for which $\hat c < 2$; this then
evolves to a trivial IR fixed point.  Thus, in all cases, the
$c$-theorem is respected.

\subsubsec{The Fate of the Scherk-Schwarz and Type 0 Tachyons: $h=1$}

Consider a torus with Scherk-Schwarz boundary conditions on the A cycle in a 
regime in which the A cycle shrinks
below string scale over some portion of the B cycle: condensing the 
tachyonic wound string breaks this handle.
This time, however, the resulting spacetime is topologically a sphere, and 
the corresponding NLSM has a gap,
flowing to $c=0$ in the IR and respecting the $c$-theorem in the strongest 
way possible for a unitary theory.

This process is intimately related to the condensation of the Type
0 tachyon: taking the radius of the A cycle to zero and
T-dualizing gives precisely Type 0, with the tower of winding
tachyons dualizing to the momentum modes of the bulk tachyon on
the non-compact dual circle. The decay of a Scherk-Schwarz torus to
a sphere thus provides a convenient model in which to locally
condense the Type 0 bulk tachyon, strongly suggesting that the
fate of Type 0 under tachyon condensation is indeed a non-critical
string theory, as has been widely conjectured.

The fate of the Type 0 tachyon has received much attention, and several 
interesting alternative conjectures
exist besides a flow to a non-critical string theory.  Perhaps the most 
fascinating is the conjecture that Type
0 is connected to Type II via condensation of a mode of the type 0 theory 
\refs{\CostaNW,\GutperleMB}\foot{The study of
localized tachyon condensation in non-compact twisted circle 
compactifications \refs{\otherRG,\GutperleBP} has also
been interpreted as supporting this conjecture; however, there is an open 
question regarding the order of limits
involved.}. In light of the $c$-theorem, as well as the decrease of the 
energy under tachyon condensation, this
conjecture requires that the world-sheet description breaks down along the 
flow if this connection between type
0 and type II is to be made by condensation of the type 0 tachyon (in 
\refs{\CostaNW,\GutperleMB} 
two modes, flux and
tachyons, were discussed). Indeed, the authors of 
\refs{\CostaNW,\GutperleMB} 
posed 
this conjecture in the context of
strongly coupled fluxbranes. It would be interesting if indeed some 
mechanism for connecting Scherk-Schwarz to
type II on a circle exists; in our context this would provide a physical 
connection between Riemann surfaces of
different spin structures.

A condensate of the zero mode tachyon in flat space does in fact source the 
dilaton (as it must, by our
discussion of the $c$-theorem above). By condensing the bulk tachyon in a 
localized fashion as described above,
however, the effect of tachyon condensation on the string coupling may be 
controlled during the perturbative,
tachyonic decay from torus to sphere (by keeping the volume of the torus 
large in string units), leading to an
NLSM on a sphere with an initially mild time dependent evolution toward 
smaller spheres. While the sphere
eventually experiences a big crunch, so that literal endstate of this system 
is necessarily nonperturbative, it
appears rather far removed from type II on a torus.

\ifig\typezip{ Tachyon condensation drives a Scherk-Schwarz torus to a 
sphere.
}{\epsfxsize2.5in\epsfbox{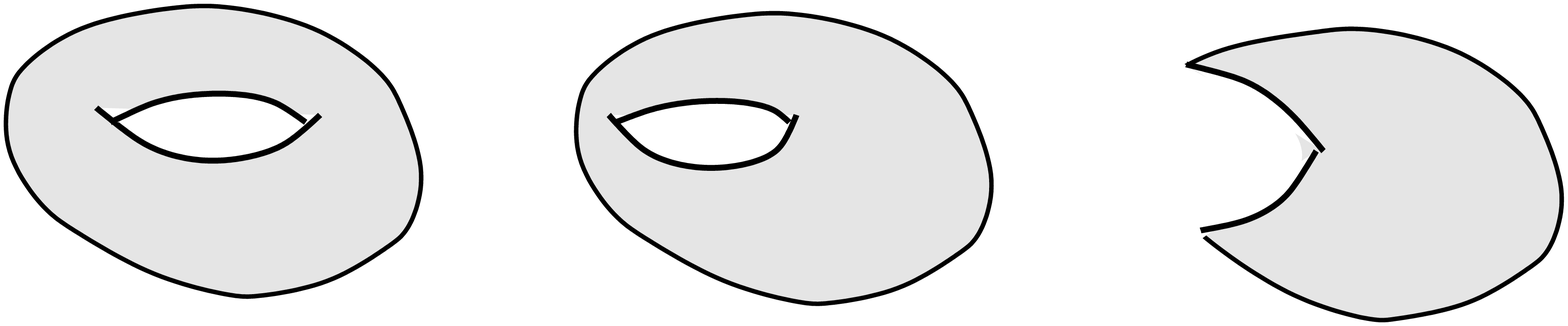}}

\subsubsec{$c$ and disconnection: $h\to h'\oplus [h-h']$}

Decays in factorization limits pose a slightly different question for the 
$c$-theorem.  Considering only the set
of degrees of freedom which localize on one component of the degenerate 
Riemann surface and repeating the above
arguments, $c$ certainly appears to be decreasing.  However, since we end 
with two decoupled perturbative
world-sheet theories, is this the correct measure? Stated differently, what 
is the correct $c$-function on the
endstate string theory with disconnected target space?

This problem is similar to the question of defining the c-function in a 
field theory with a double well
potential, working at energy scales below the barrier height.  (As we will 
see in the next section, this is a
good model for our system in fact.) Equivalently, we may consider the 
scaling of the entropy of free fields as a
function of temperature in such a double well configuration.  The c-function 
decreases consistently in the full
quantum field theory.  Of course, this does not have operational 
observational meaning in the causally
disconnected regions, but is a consistency check on the analysis of the 
quantum field theory on the world-sheet.

\newsec{Spacetime Physics of the Transition}

We start this section by analyzing the
effects of the time-dependent perturbation on propagation
of modes through the tube, which builds on the RG results just
reviewed and applies to the process at the level of string
perturbation theory. We then make some comments on the
time-dependent process in the full theory, particularly
emphasizing the role of particle and string production in both
topology-changing and tachyon decay processes.

\subsec{Transmission Barrier}

In this subsection, we use the nonlinear sigma model to study the
effect of winding tachyon condensation on the propagation of the
string world-sheet into the tube.

Consider a tube coordinatized by $\theta$ and $r$, with radius in the $\theta$ direction reaching a sub-string
scale minimum at a point $r=r_0$ (or equivalently the radius of the T-dual variable $\tilde\theta$ reaching a
maximum at $r_0$). We would like to consider the scattering of modes going into the tube from large negative $r$
to see if they get transmitted through the tube after tachyon condensation. If our conjecture about the
topology-changing process is correct, the waves should be completely reflected back to large negative $r$ by the
end of the process.  We will check that the tachyon condensation introduces a barrier to penetration through the
tube.

Including the time-dependence, the tachyon vertex operator is
\eqn\Ttime{\int d^2\sigma d^2\theta ~T(X)}
with $T(X)=\hat T(R) e^{\kappa X^0}\cos[w\tilde\Theta]$.
Because of the weak curvature, the $r$-dependence in the
lowest-lying tachyon vertex operator is mild.  It decays to zero
exponentially for values of $r$ outside the region where the
winding mode is tachyonic and rises to a maximum at $r=r_0$, where
the tube radius reaches its minimum. More generally, we can
consider momentum modes in the $r$ direction, producing
oscillation at a scale $k_r$ on top of the mild exponential
falloff of the wavefunction.

In components, this leads to a bosonic potential in the
world-sheet matter theory which is of the form
\eqn\Ttimebos{U[X]=\del_\mu T\del^\mu T= \biggl\{(-\kappa^2
\cos^2[w\tilde\theta]+w^2\sin^2[w\tilde\theta]) \hat
T(r)^2+{(\del_r\hat T)^2} \cos^2[w\tilde\theta]\biggr\}e^{2\kappa X^0} }
The kinetic terms include a mild $r$-dependent metric
\eqn\kinT{{\cal L}_{kin}\sim -\del_aX^0\del^a X^0+ g_{\tilde\theta\tilde\theta}
(r)\del_a\tilde\theta\del^a\tilde\theta + g_{rr}\del_ar\del^ar} .

Upon condensing the tachyon, the world-sheet theory becomes a
nontrivial sigma model on a time-dependent target space, subject
to the constraints of two-dimensional (1,1) supergravity.  The
essential classical effects of supergravity in the world-sheet
theory are as follows.  As discussed in \eg\ \DaCunhaFM, $X^0$,
with its negative kinetic term, can be traded for the conformal
factor of the $2d$ metric; its action arises from a conformal
transformation of the form $\exp{(\gamma X^0)}$ in the
gravitational action on the world-sheet.\foot{Given the time
dependence in the ambient dilaton and volume from the tadpoles in
\poten, the action should be supplemented by a linear dilaton
term, \ie\ a conformal coupling proportional to $\int d^2\sigma
R[e^{ \gamma X^0} \hat g] (2h-2)\Phi'(X^0)$ where $\Phi$ is the
log of the effective string coupling and $\hat g$ is the fiducial
metric from gauge fixing which we take to be flat.} The negative
term in the potential energy \Ttimebos\ arises from supergravity
in this sense--it came from the derivative of the superspace
potential $T(X)$ by $X_0$ -- and is reminiscent of similar
negative terms in the scalar potential arising in supergravity in
higher dimensions.

The potential \Ttimebos\ exhibits a barrier to penetration into
the region where the tachyon vertex operator has support. We
consider a regime of parameters
\eqn\ineqk{\kappa^2<k_r^2\ll w^2 .}
The first inequality ensures that the momentum $k_r$ in the $r$
direction is sufficient to overcompensate the negative $-\kappa^2
\cos^2[w\tilde\theta] $ term for some range of $r$, so that these
terms in combination with the leading term proportional to $w^2
\sin^2[w\tilde\theta]$ provide a classical barrier for all values
of $\tilde\theta$. The second inequality ensures that the leading
interaction term for $\tilde\theta$ is the supersymmetric
sine-Gordon interaction, a fact that will facilitate analysis of
the quantum theory.

In particular, although the action exhibits a classical potential barrier for the string in the region where the
tachyon exists, in general one needs to include quantum corrections in order to determine the net effect of the
interaction terms on modes impinging on the tube.  We can use the known results for the RG flow in the model to
analyze this, as follows.

First, for modes for which $r$ depends weakly on the world-sheet
coordinates, \ie\ modes with small momentum in the $r$ direction
and no oscillator excitations, this barrier indeed survives in the
quantum theory. This is because the $r$-dependent factors in
\Ttimebos\ behave to good approximation like coupling constants
multiplying $\tilde\Theta$-dependent operators. Since we required
\ineqk\ so that the $w^2 \sin^2[w\tilde\theta]$ term dominates,
the long distance physics of the perturbed sigma model is the long
distance physics of the supersymmetric sine-Gordon theory, which
has a massive spectrum as we reviewed in the last section.  At low
energies in the $r$ sector, the massive $\tilde\Theta$ excitations
will not be generated. Putting the $\tilde\Theta$ sector in its
vacuum then yields a surviving potential for $r$ in the action
\Ttimebos, which blocks low-energy modes from getting through the
tube (ignoring tunneling effects at very weak string coupling).
Thus for low-energy modes, this leading-order contribution in the
tachyon is sufficient to block penetration. Moreover, after
integrating out the $\Theta$ sector the remaining interactions for
$r$ are themselves tachyonic perturbations, whose effects grow in
the infrared.

Secondly, one can use the same procedure for more general modes,
as long as the masses in the $\tilde\Theta$ sector exceed the
scale of fluctuation in the $r$ sector.  As one takes the limit of
large tachyon vev in \Ttimebos, the $\tilde\Theta$ masses
increase, as can be seen from the S-matrix results and direct
calculations of soliton masses in
\refs{\ShankarCM,\AhnUQ,\BajnokDK}. Thus as we increase the
tachyon perturbation, the barrier affects more energetic modes. It
is reasonable to expect that a limit exists in which all modes are
blocked.

Note also that the $(\del_r T)^2$ term introduces at least one
local minimum in $r$ (and generally more depending on $k_r$) in
the middle of the tube in addition to the potential barrier. This
may reflect the extra vacua predicted by the Witten index.

Finally, in order to define the world-sheet path integral, it may
be useful to continue to Euclidean signature in the target space,
setting $X^0\equiv i t_E$.  In doing this, we must consider
$\cosh[\kappa X^0]=\cos[\kappa t_E] $ instead of the pure
exponential, which is reminiscent of spacebrane constructions.
This approach commits us to studying amplitudes in the Euclidean
vacuum, which is a reasonable choice.  From this we obtain
positive kinetic terms and a bosonic potential
\eqn\TEuclbos{\eqalign{U_{{\rm bosonic}}= & \hat T(r)^2\bigl(\kappa^2
\sin^2[\kappa
t_E]
\cos^2[w\tilde\theta]+w^2\cos^2[\kappa t_E]\sin^2[w\tilde\theta]\bigr) \cr &+{(\del_r\hat T)^2}
\cos^2[w\tilde\theta]\cos^2[\kappa t_E]\cr }}
Like \Ttimebos, this exhibits a potential barrier and a similar
analysis to that above applies.

\subsec{Comments on the time-dependent process}

The renormalization group results just reviewed in \S2\ and
applied in the last subsection provide strong evidence for
connections between different spacetime topologies in the
configuration space of string theory. Such a connection was
established on the moduli space in Calabi-Yau target spaces in
\refs{\classtop,\quantop}. 
In the quantum theory at nonzero string coupling,
the time-dependent dynamics of topology
change processes and tachyon decay processes are very rich, with
particle, string, and brane production effects playing a crucial
role. Before describing this process in our case, let us briefly
discuss the topology-changing processes studied previously as well
as earlier examples of tachyon decay processes.

In highly supersymmetric situations where the moduli space has
been shown to contain configurations of different topology,
particle production effects play an important role in limiting the
extent to which topology-changing processes occur dynamically. For
example, in the conifold transition in ${\cal N}=2$ vacua of type
II string theory \quantop, the low-energy effective description of
the transition is as a transition between the Higgs and Coulomb
branch of an ${\cal N}=2$ supersymmetric field theory \stromcon\
(a realization which resolved a longstanding puzzle about the
behavior of the string amplitudes near this point). Hence one
might attempt to effect the change of topology dynamically by
rolling the scalar fields toward the point where one branch joins
another, for example rolling on the Coulomb branch toward the
point where a hypermultiplet becomes massless. However, because
the mass of the hypermultiplet is decreasing toward zero along
this trajectory, the moduli space approximation breaks down badly,
regardless of how small one takes the initial scalar field
velocity; the nonadiabaticity parameter is of order $\dot
m/m^2\sim \dot\phi/\phi^2$ where $m=\phi$ is the mass of the light
hypermultiplet in terms of the canonically-normalized scalar field
$\phi$. Light hypermultiplets are created, and their energy
density back-reacts on the motion of the Coulomb branch scalar in
a simple, calculable manner.  This results in rapid trapping of
the scalar field at the intersection point between the two
branches \trapping, rather than dynamical evolution toward a large
radius space with different topology. In less symmetric
situations, potential energy and cosmological evolution may
dominate over these kinetic effects and yield dynamical
topology-changing transitions.  In the case of flop transitions,
there are no light states involved in the stringy resolution of
the topology-changing transition, so these may occur dynamically
even in cases with extended supersymmetry, perhaps as in
\GreeneYB.

In earlier studies of tachyon decay processes, the time-dependent
problem at nonzero string coupling involved the production of many
excitations, including a gas of localized D-branes as well as
strings and gravity modes \tachtime. In a technical analysis, this
production can be avoided by considering a strict limit of zero
coupling \refs{\sen,\AdamsSV}, but it is expected to occur in the
physical problem, at nonzero coupling.

In our case, the dynamical process involves both effects just reviewed.  The beginning of the process consists
of rolling in complex structure moduli space $\tau$ toward a point with a small handle or toward a factorization
limit\foot{Note that as mentioned above there is a 1-loop contribution to the scalar potential driving the
Scherk-Schwarz circle to shrink, so this is a natural process to consider in the absence of metastabilizing
fluxes.}. Again, the masses of other modes depend on $\tau$, and those which become light (such as the winding
modes $T$ around the tube) get produced. This traps $\tau$ in the region where $T$ has mass squared less than or
equal to zero. Once in this region, the tachyon condenses, producing many excitations from the energy liberated
in the process, as in \tachtime.

It is worth noting that the back reaction effects of the decay products may be mitigated by a more elaborate
setting, as follows. We could embed our setup in the class of de Sitter models discussed in \SaltmanJH, while
relaxing the flux contributions to allow for the small handle to develop.  If the resulting system after the
tachyon condensation is in a basin of attraction of one of the de Sitter solutions \SaltmanJH, then the late
time de Sitter expansion will dilute the decay products of the tachyon condensation process.

\newsec{Charges, D-branes, and Fluxes}

In this section, we address the fate of charges in the system, and
generalize our analysis to situations with fluxes turned on.

\subsec{Fundamental String Charges}

When a handle decays, the conserved charges associated with it are
also lost in the process.    By condensing {\it one} winding mode
to annihilate the handle, we have destroyed {\it two} cycles of
the Riemann surface, namely the $a$-cycle $A$ on which the winding
string is wrapped, and its intersection-dual $b$-cycle $B$. Each
cycle corresponds to a conserved fundamental string winding
charge, whose gauge field in the eight-dimensional effective
theory we will refer to as $F_A$ and $F_B$ respectively.

The effective action for these field strengths is (by a simple application of the analysis in \S2\ of
\SaltmanJH, starting from the 3-form field-strength $H$ and decomposing it in a basis of 1-forms on the surface
to reduce to the 2-form field strengths $F$ pertaining to the gauge fields in the $8d$ effective theory)
\eqn\couplinggeneral{\int_{8d} \int_\Sigma H \wedge \ast H = \int_{8d} {\cal A}_{jk}(\tau) F^j\wedge_8 \ast_8
F^k. }
Here $j,k$ index the quantum numbers on the $a$ and $b$ cycles of
the Riemann surface, and ${\cal A}_{jk}(\tau)$ is the $2h\times
2h$ matrix
\eqn\Amatrix{{\cal A}(\tau) =  i \pmatrix{ 2 \tau (\tau-\bar\tau)^{-1} \bar\tau &
    -(\tau + \bar\tau)(\tau-\bar\tau)^{-1} \cr
                -(\tau-\bar\tau)^{-1} (\tau + \bar\tau) & 2 (\tau-\bar\tau)^{-1}}}
where $\tau$ is the $h \times h$ period matrix of the Riemann surface. This coupling reduces at genus 1 to
\eqn\couplingfunction{\int d^8 x\sqrt{g} {1\over\tau_2^2} | F_A + \tau F_B|^2 }
where $A$ and $B$ are the two one-cycles of the surface. The
formula \couplingfunction\ is also a good approximation at higher
genus in situations where the handles not participating in the
process are nearly decoupled from the $a$-cycle $A$ on which the
winding tachyon is wrapped and its dual $b$-cycle $B$. As we
evolve well into the tachyonic regime, $\tau_2$ becomes small, so
$F_A$ is weakly coupled while $F_B$ becomes strongly coupled.

The condensing tachyon, since it is a string wound around the
cycle $A$, is charged under the $A$-cycle gauge field. This gauge
field is therefore massed up by the Higgs mechanism. The fate of
the gauge field coming from the $B$-cycle is more mysterious --
the strong coupling suggests that it confines classically
\KogutSN.  This puzzle and its resolution have been seen in the
open string tachyon problem \openconfine as well.
\ifig\breakage{
A pictorial representation of the fate
of the B-cycle winding charge.
If there are $N$ strings on cycle $A$,
with indefinite $N$, then
a string on cycle $B$ can break.
}{\epsfxsize2.0in\epsfbox{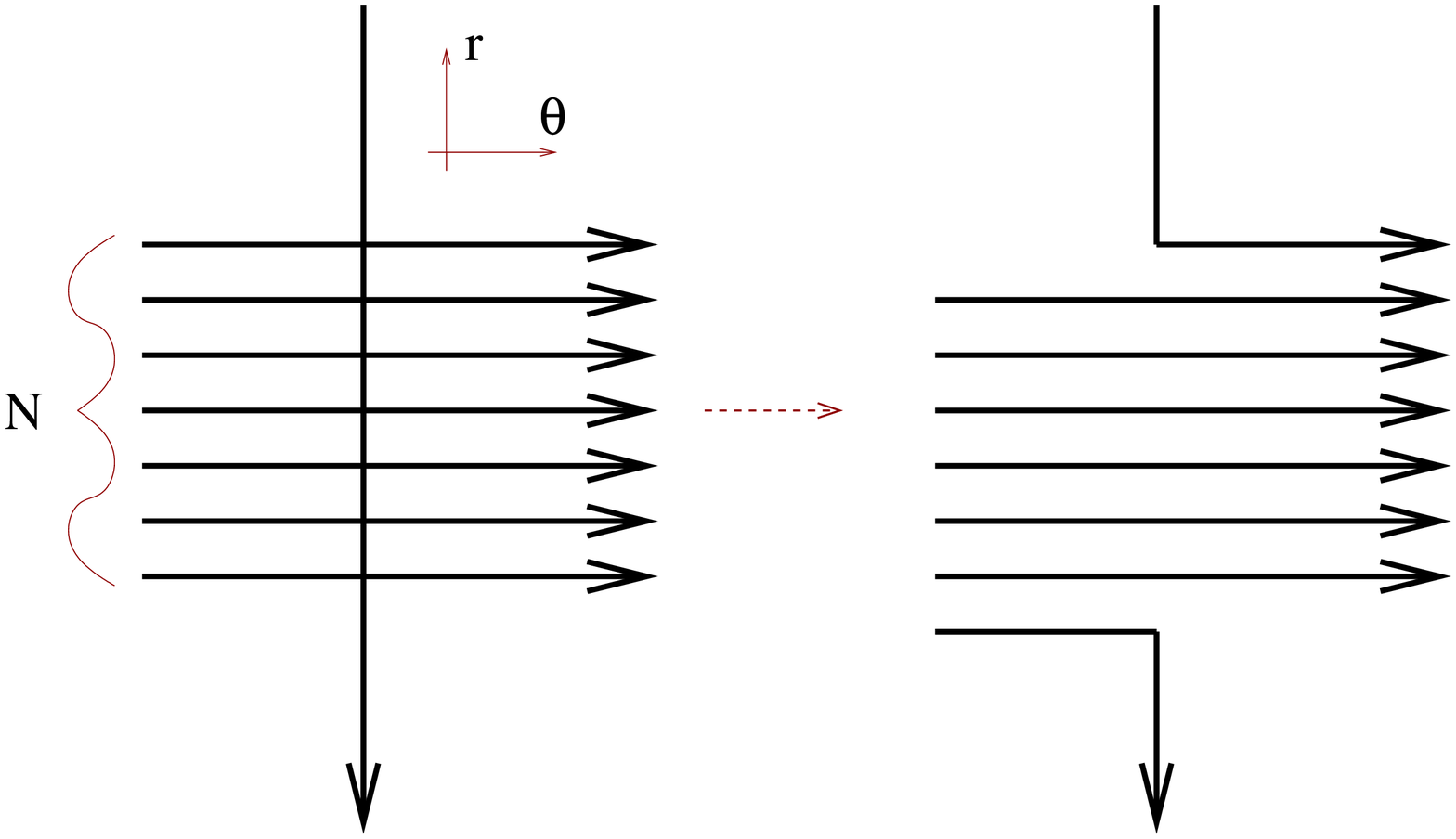}}

As another probe of the fate of the $B$-cycle gauge group, consider, before the condensation, a state with a
string wound on the $B$-cycle at some point $x$ in the remaining 7 spatial dimensions and a string wound with
the opposite orientation at a point $y$. From the 8-dimensional point of view, this is a charge-anticharge pair.
Since these charges are codimension seven, they cannot affect the condensation process globally. Nevertheless,
as the $A$-cycle tachyon condenses, and the effective $B$-cycle coupling grows stronger, the lines of $F_B$-flux
running from one charge to the other will collapse into a tensionful flux string \KogutSN. These lines of flux
may temporarily prevent the handle from collapsing. However, the charges will experience a strong attraction and
annihilate as soon as possible, allowing the handle to decay subsequently.

\ifig\confimenentfigure{ A geometric realization of classical confinement. Each point in this picture represents
a circle homotopic to the $A$-cycle. $y$ is the direction in the remaining 7 spatial dimensions in which the
charges are separated. The dashed lines represent $H_{{\rm NS}}$-flux lines between the fundamental string
sources. }{\epsfxsize3.0in\epsfbox{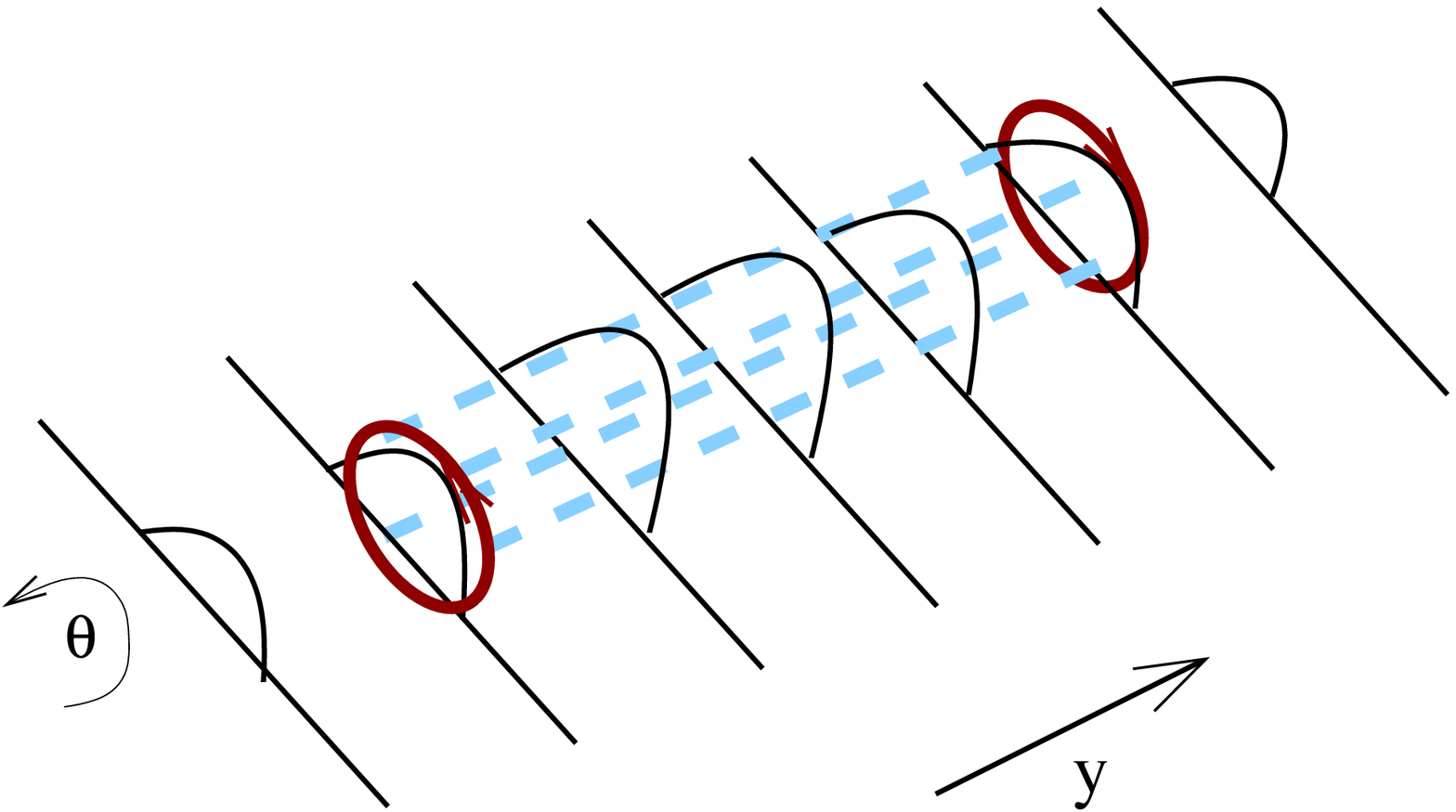}}

On a general Riemann surface, there are other cycles not directly
participating in the decay process occurring on the $A_1-B_1$
handle.  For those handles with antiperiodic boundary conditions
for fermions on one or both cycles, one might wonder if their
decays are induced by the decay process occurring on the first
handle. In the full time-dependent system, in the absence of
fluxes or other ingredients to metastabilize, the allowed decays
will all occur in time. At the classical level however, one can
say more.  The strings wound around these other cycles do not
couple linearly to the tachyon wrapped around the $A_1$-cycle, so
at the classical level their condensation (and concomitant Higgs
mechanism) will not be induced.  Their gauge couplings are also
not getting strong as in \couplingfunction, so we do not expect
confinement to set in. From the world-sheet point of view, the
global symmetries associated with the winding currents for these
other cycles should be unaffected by the destruction of the
$A_1-B_1$ handle.

\subsec{D-branes}

In type II theories, D-brane charges also disappear in the tachyon
decay process.  These charges are analagous to the $B$-cycle
fundamental string charge described in the previous subsection:
the gauge groups in question are not Higgsed by the fundamental
winding string condensate on the $A$-cycle, but could be Higgsed
by a condensate of a heavy non-perturbative object later in the
time-dependent process.

In previous studies of tachyon condensation processes, D-brane
probes yielded useful independent information \AdamsSV; in the
present context this does not happen because the initial flow
occurs when the geometry is not singular, as it was in \AdamsSV.

\subsec{Flux Backgrounds}

We can also consider this process in a situation with mild fluxes
on the Riemann surface.  As long as these are sufficiently weak,
the region of complex structure moduli space with small handles
remains accessible.  Suppose, for example, that we have 1-form
flux $F$ on the compact cycle $\gamma_\theta$ of the tube (where
$F$ could come from a $p$-form field strength in higher
dimensions, with one leg on the Riemann surface in question):
\eqn\tubeflux{\int_{\gamma_\theta} F = Q}
When the tube caps off in the tachyon decay process, the integral
\tubeflux\ still holds in the remaining region, so a source must
appear. In particular, a brane charged under $F$ must appear at
the tip of one cap while the corresponding antibrane appears at
the tip of the other.  Hence, in the presence of flux, this
process produces brane-antibrane pairs as well as change of
topology.

\newsec{Discussion}

We have shown that a tachyon decay process in which a localized
Scherk-Schwarz tachyon decays away the central charge in its
region leads to topology change processes in which handles decay
away and Riemann surfaces break into separate components. Both are
expected from spacetime energetics \poten\ and follow from
world-sheet renormalization group arguments.  The effect is likely
to be very efficient dynamically as it is driven by a tachyonic
mode rather than a massless modulus field.

It is interesting to consider the possibility of a dual gauge
theory describing the local physics of the tube, perhaps via some
embedding of the system in AdS/CFT. In such a situation, we would
expect the winding string to be dual to a Wilson loop operator.
Condensation of this Wilson loop would then imply that the cycle
on which the string is wrapped is contractible, since it can be
filled in with a string world-sheet. Considering the winding
circle as Euclidean time, this is the same argument that shows
that a vev for the Polyakov loop implies that the dual geometry
contains a black hole horizon \AharonySX. This agrees with our
'capping-off' picture. It would be interesting to understand such
a relationship in detail.

The change in the number of components of the Riemann surface,
suggesting formation of baby universes perturbatively via stringy
physics, is particularly striking.
Previous work on baby universe creation involves
quantum-gravitational tunneling computations
(see \eg\ \LindeSK\ and references therein). The separation of
components we see here from a {\it perturbative} instability is
quite different,
and does not depend on subtleties involved in
Euclidean quantum gravity computations.\foot{ At {\it large}
radius for the Scherk-Schwarz tube, the Witten ``bubble of
nothing" \WittenGJ\ may play a similar role. However, this
solution is subdominant to the perturbative dynamics, including
higher order corrections driving the complex moduli toward the
tachyonic regime in situations without stabilizing fluxes.  In
situations with stabilizing fluxes, the decay of fluxes by brane
nucleation appears to be the leading decay mode so it is not clear
if and when the bubble of nothing dominates.} As such, our results
provide simpler motivation for the need to make sense of the
observables in situations where space separates into distinct
components. 
(See for example \wormholerefs, though here the baby
universe is not necessarily Planck sized as in \wormholerefs; the
`dust' vacua may however play this role.).\foot{Of course, other
solutions exist with causally disconnected regions, such as the
connected discretuum of metastable de Sitter solutions in string
theory, whose resolution may be similar.}

Given the dramatic nature of this effect, let us
note two conceivable loopholes that could evade this conclusion (neither of
which
appears plausible):  (1)  perhaps the effect takes infinite time according
to the appropriate
observer, or (2) perhaps there is some very attractive endpoint for the RG
flow in the $\tilde\Theta$
sector which is a nontrivial IR stable fixed point rather than the trivial
IR limit
with a mass gap that emerges from
all analyses of the model to date.  Realizing loophole (2) would
require some extremely surprising RG behavior,
while (1) would require some mechanism for slowing down the
transfer of energy from potential energy \poten\ to decay products
via the channel we have identified.  Some of our arguments (such
as the world-sheet $c$-theorem) depend on a small string coupling,
something we can tune to obtain control 
as in \refs{\AdamsSV,\sen}.  The
basic spacetime energetics of the process (in particular the
reduction of the 8d potential energy by the condensation of $T$)
appears robust in the presence of interactions, though the
possibility of a `remnant' alternative could in principle arise
via as yet unknown strong coupling effects.

In order for a truly separated baby universe to form, the process
we described must take place everywhere in the remaining eight
dimensions. It may be more natural in early universe cosmology for
the complex structure moduli to be different in different regions,
and hence only experience a given topology-changing process
locally. However, if it happens the same way within a given
horizon volume, the effect is the same as far as any given
observer is concerned.

In either case, the resulting dynamics of the effective 8d effective field theory appears remarkable: the tachyon
perturbation must 
be such that the spectrum decouples into two sets of observables whose interactions
vanish - including the appearance of a new spin-2 state whose longitudinal mode must decouple at the endstate.
Various authors have considered such processes within the context of effective field theory
(see \eg\ \ArkaniHamedSP); it would be fascinating to understand this process in the stringy effective field theory.

The tachyonic topology-changing dynamics we identified here also
suggest the possibility of 
{\it topological} topological defects.  In
particular, the type II theory in the backgrounds we have focused
on has a complex tachyon, which could yield cosmic strings whose
cores have different topology in the internal dimensions.

In addition to sharpening the conceptual puzzle of how to treat
causally disconnected regions in gravity, these considerations
point to new scenarios for string cosmology.  For example, the
complex structure modulus, with appropriately tuned potential,
combined with the tachyon, provide an analogue of hybrid
inflation, similar to brane-antibrane inflation but in the closed
string sector.  More generally, it will be interesting to
investigate whether the existence of one-cycles in the early
universe (a generic feature given stringy energy densities at
early times) and their subsequent decay may produce interesting
signatures.  In the context of low-scale gravity scenarios it is
also interesting to ask about signatures of decay (and formation)
of handles.

One could also consider a configuration in which a sphere bubbles off
of a surface of any genus, starting from a configuration with a locally
pinched neck.  As in the factorization case, this automatically
has antiperiodic boundary conditions for fermions.  However, the starting
pinched configuration required is not a stable endpoint of the Ricci
flow in this case, and hence 
this pinched configuration is separated from 
the constant-curvature configurations by a potential barrier.
It may be relevant in the early universe; high temperature effects could
allow the system to fluctuate over the barrier.

Finally, 
on rather more speculative ground, 
the fact that these
topology-changing processes involve the removal (or appearance, in
principle) of handles 
though tachyon condensation
suggests a host of intriguing generalizations.  For
example, since string theory is not only a theory of
strings, it is natural to wonder whether a $p$-brane wrapping a
sufficiently small 
non-supersymmetric
cycle might go tachyonic 
(as suggested in \openconfine),
and
mediate topology change through the removal of the shrinking
$p+1$-handle.  
This would provide a lovely link between handlebody
theory and string theory.  
Mathematicians \ricciflow\ have studied RG 
flows in target spaces of various dimensions
at the level of the leading order in $\alpha'$.  
They have been motivated to
perform surgery manually, `capping off' their spaces 
at a 
cutoff length scale in a way very reminiscent 
of the process we have described.  It would
be very interesting to see if string or brane theoretic tachyons could 
effect this surgery naturally.

\bigskip
\centerline{\bf{Acknowledgements}}

We would like to thank Mohsen~Alishahiha, Nima~Arkani-Hamed, Savas~Dimopoulos, Steve~Giddings, Matthew~Headrick, Albion~Lawrence,
Alex~Maloney, Shahin~Sheikh-Jabbari, Larus~Thorlacius, David Tong and especially Shamit~Kachru and Steve~Shenker
for discussions. A.A. is supported by a Junior Fellowship from the Harvard Society of Fellows. The rest of the
authors are supported in part by the DOE under contract DE-AC03-76SF00515 and by the NSF under contract 9870115.
E.S. thanks the IPM and ISS String School in Iran and the PIMS and PNW String Seminar in Vancouver for
hospitality during parts of this work.

\appendix{A}{Linear Sigma Model description}

In this appendix, we describe the effects of the winding tachyon using linear sigma models.  That is, we
consider a field theory model whose low-energy limit describes the 
physical problem of interest, following
a method developed in \WittenYC\ in Calabi-Yau compactifications 
and applied 
to previous
tachyon decay problems \AdamsSV\ in, for example \refs{\VafaRA,\chicago}.
We focus on a noncompact region of the Riemann surface containing a single handle, and
construct a linear model with a low-energy configuration space that flows from a one-sheeted hyperboloid to a
two-sheeted one (see the figure below), exhibiting the local topology-changing process of interest via a description in
which the relevant operator corresponding to the tachyon is realized in a simple way. This method will apply to
the renormalization group flow aspect of the problem, discussed in \S2. It will also make manifest the extra
vacua required by conservation of the Witten index, and will clarify the fate of the GSO projection in the flow.

Relative to the case \refs{\AdamsSV,\VafaRA,\chicago}, there is a
complication regarding the level of supersymmetry in the problem.
As we discussed below \TsupvertII, the relevant operator arising
from the tachyon vertex operator does not preserve ${\cal N}=2$
supersymmetry unless the smallest circle in the tube is zero size.
Hence, in order to describe the process starting from a smoother
space, we ultimately consider a model containing soft ${\cal N}=1$
deformations away from a $(2,2)$ supersymmetric gauged linear
sigma model (GLSM).  We analyze this model using its
superrenormalizability and its reduction to $(2,2)$ in an
appropriate limit. The deep IR regime of the model contains
nontrivial corrections to potential energy terms, but we find we
can tune these effects away to sufficient accuracy for our
purposes. It would be interesting to pursue the physics of this
model in more detail than we have space for here.

Before moving to our ultimate construction, it will be useful to
keep in mind the following simple ${\cal N}=1$ model which
contains some of the essential features that will be encoded in
our larger model.
\doublefig\caps{ The vacuum manifold a) at $\rho < 0$; b) at $\rho
> 0$. }{\epsfxsize2.0in\epsfbox{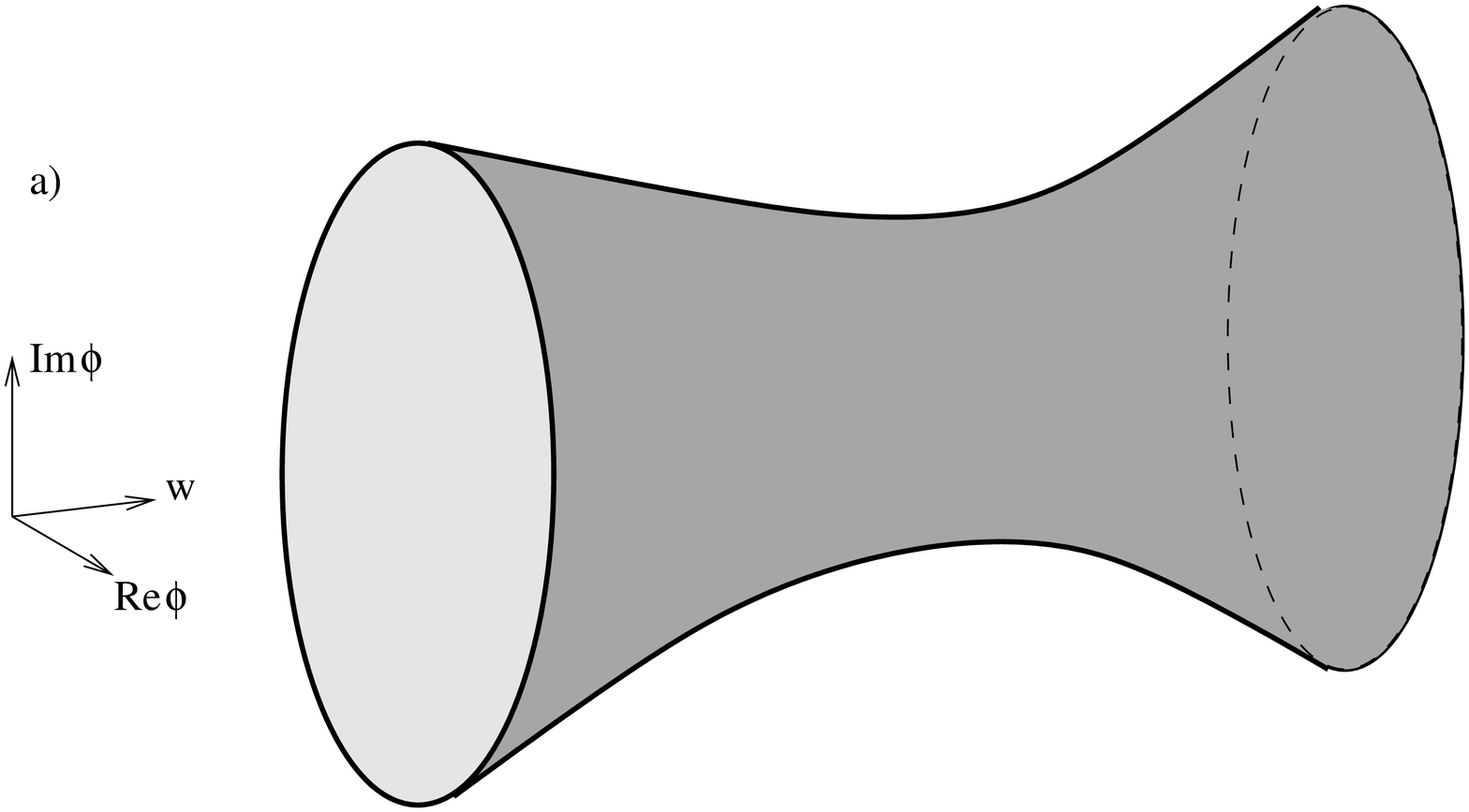}}
{\epsfxsize2.0in\epsfbox{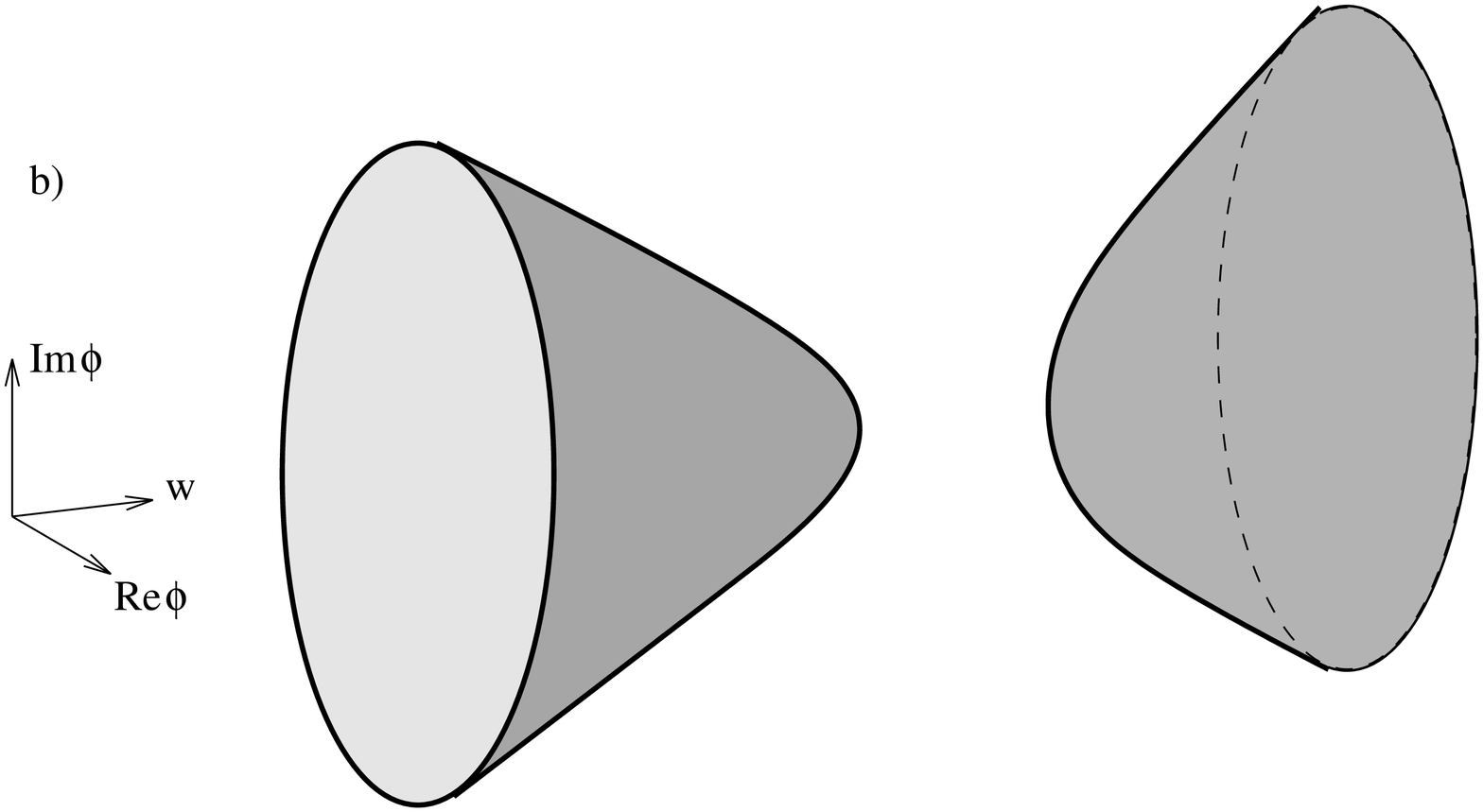}}
The desired geometry looks like the locus
\eqn\desired{
 w^2 - |\phi|^2 = \rho  , ~~  w \in \IR, \phi \in \IC. }
As $\rho$ goes from negative to positive, the manifold changes
from a one-sheeted hyperboloid to a two-sheeted one, as in 
the figure.
A very simple (1,1) model with this vacuum manifold in the
low-energy limit has real multiplets $w, \sigma$ and a complex
multiplet $\phi$, with potential terms
$$
\int d \theta_+ d \theta_-
~ \sigma \left(  w^2 - |\phi|^2 - \rho  \right) ;
$$
here $\theta_\pm$ are left- and right-moving coordinates on a $(1,1)$ superspace.  For our purposes we will need
a model (to be elaborated below) for which this geometry arises at low energies, including quantum corrections.
Quantum corrections also determine the direction of flow of $\rho$, and we will find that our model predicts the
appropriate direction, with $\rho$ flowing toward more positive values.
Before moving on, we can use this model (in which many of the NLSM
couplings are hidden in $\rho$) to see that the coupling $\rho$
multiplies a winding operator, as follows.


Consider T-dualizing in the direction of ${\rm Arg}(\phi) =
\theta$, which goes around the hyperboloid. This is accomplished
\refs{\BuscherQJ, \RocekPS} by introducing a gauge field $F = dA$
under whose gauge transformation $\theta$ shifts, and whose theta
angle is an extra field $\tilde \theta$,
\eqn\buscheraction{ S =  \int \left( (\del \theta + A)^2 +
 \tilde \theta F +{1\over e^2}F^2 \right) .}
This dynamical theta
angle $\tilde \theta$ acquires kinetic terms upon integrating out
the gauge fields, and becomes the T-dual coordinate.\foot{
The kinetic term we have included in \buscheraction\ 
does not affect the analysis of 
the T-duality below the high scale $e$.}
With this in
mind, the coupling $\rho$, which in (2,2) models is the
Fayet-Iliopoulos (FI) parameter, turns on the winding tachyon
because it controls the action of a vortex of the gauge field and
scalars. In a vortex configuration, the support of $F$ is
delta-function localized near the centers of the vortices. So, in
a background of $\tilde \theta$, a vortex at a point $z$ on the
world-sheet contributes \eqn\vortexcontribution{ e^{ - S_{{\rm
cl}}(\rho) } e^{ i \tilde \theta (z)}, } where $S_{{\rm
cl}}(\rho)$ is the $\rho$-dependent vortex action.
\vortexcontribution\ is an insertion of the winding vertex
operator with a $\rho$-dependent chemical potential, $T =
T(\rho)$. The sum over such vortices automatically reproduces the
coulomb gas expansion in the winding deformation.

This is elegantly packaged in the (2,2) {\it gauged} linear sigma
model (GLSM). Here the gauge field is present from the beginning,
and its vortices can be BPS. In the $(2,2)$ case
\refs{\MorrisonYH, \HoriKT}, the T-dual direction couples as a
dynamical theta angle to the GLSM gauge field. The coupling of the
winding operator is complexified to $ e^{ \rho + i \theta_G}$,
where $\theta_G$ is the theta-angle of the GLSM gauge field. This
claim is closely related to the fact that in the Calabi-Yau case
\WittenYC\ the vortex sum reproduces the sum over world-sheet
instantons of the nonlinear sigma model.

It is for this reason that we obtain our hyperboloid geometry from
an equation of the form \desired\ following from a $(2,2)$ D-term.
Another possible starting point would be to obtain the initial
hyperboloid from a (2,2) F term via a superpotential $ W = P (\phi
\eta - \mu ) $ where $\mu$ is a constant, and $P, \phi, \eta$ are
chiral fields. The vacuum manifold of this model is the 'deformed'
one-sheeted hyperboloid $\phi \eta = \mu$. However, the tachyon is
still a solitonic winding operator so the linear sigma model
description would not provide a simple description of the flow.


Indeed, there is a good precedent for
the FI parameter playing the role
of the vev of a localized (winding) tachyon
\VafaRA.
In these $\IC^n/\IZ_N$ models, the twisted tachyons arise from
strings which stretch from a point to its image under the
orbifold; if these strings from the $k$th twisted sector are
pulled away from the tip of the cone, they are forced to wind
around it $k$ times. In the limit $N\to \infty$, this $k$ becomes
an integer-valued winding charge. The models we are studying,
therefore, can be considered to promote this `fractional'
localized winding tachyon to an honest winding tachyon. It is
therefore natural that we should find models which realize a
condensation whose behavior is not dissimilar to that of \VafaRA.


\subsec{Stringy (2,2) model}

\def\etap{\eta_+}
\def\phip{\phi_+}
\def\phitwo{\phi_{-2}}
\def\etatwo{P_{-2}}

With this motivation, we will start by considering the following
(2,2) gauged linear sigma model, following the conventions in
\WittenYC. Consider the GLSM with one $U(1)$ under which the
D-term is
$$ D = |\phip|^2 + |\etap|^2 - 2 |\phitwo|^2 - 2 |\etatwo|^2 - \rho .$$
The subscripts of the fields label their
charges under this $U(1)$.
We will add the superpotential
$$ W = m \etatwo \phip \etap .$$
Here $m$ is a scale which sets the masses of the fluctuations
transverse to the vacuum manifold of the F-terms, and for
convenience can be considered equal to the scale $e$ of the gauge
coupling. The coupling $\rho$ flows logarithmically to $+ \infty$
in the IR \WittenYC\ because the sum of the charges is
$$ Q_{{\rm T}}  \equiv \sum_i Q_i = - 2 .$$
There is an $SU(2)$ symmetry acting on
$ ( \phip, \etap)$ as a doublet.
The F-term equations are
\eqn\ftermsofnubbin{
0 = F_{\etatwo} = m \phip \etap, ~~~
0 = F_{\phip} = m \etatwo \etap, ~~~
0 = F_{\etap} = m \etatwo \phip.
}

We now quickly analyze the vacuum manifold in the two
semiclassical regions where $ |\rho| $ is large. This will
determine the target space of the effective nonlinear sigma model
below the scales $m, e$. When $\rho \to - \infty$, either
$\phitwo$ or $\etatwo$ must be nonzero on the vacuum manifold. If
$\etatwo =0$, the remaining F-terms give $\phip \etap = 0$.  This
is an apparently singular hyperboloid
with two branches $\phip\neq 0, \etap\neq 0$. If $\etatwo \neq 0 $, both
$\phip$ and $\etap$ must vanish. If both $\phip$ and $\etap$
vanish, $\etatwo, \phitwo$ are unconstrained and parameterize a
$\IP^1$.
There is a residual $\IZ_2$ of the gauge group which is unbroken on this branch. The three branches intersect at
$\phip=\etap=\etatwo=0$. Note that for generic $\rho$ the model is not actually singular since the extra branch
is compact. Further, as long as $\theta_G \neq 0$, the model is nonsingular for any $\rho$.  

It may be interesting to consider in more detail the singularity
at $\rho=0=\theta_G$.  In Calabi-Yau models \WittenYC, the analogous
singularity reflects the presence of light non-perturbative objects 
in this limit \stromcon.  In our case, it may reflect light
wrapped D-strings, whose condensation as discussed in \S4.2\ would provide
a natural mechanism for confinement of RR charges in the process
along the lines of \openconfine.  

When $\rho \to + \infty$, $\phip$ and $\etap$ cannot
simultaneously vanish.  The F-terms therefore set $\etatwo =0$.
Further, the two branches of $\phip \etap =0$ are disconnected. On
the $\phip \neq 0$ branch, the D-term equation is of the form
\eqn\Dtermoncap{ |\phip|^2 - 2 |\phitwo|^2 = \rho . } The $U(1)$
gauge symmetry can be fixed to set $\phip$ real and positive. This
is a single `cap', with a negative curvature related to $\rho$. An
identical analysis applies to the branch with $\etap$ nonzero,
which gives a second `cap' disconnected from the first.

In this model, therefore, the physics of the condensed phase is exactly as desired, and depicted in \caps. The
starting point, however, is somewhat obscured by string-scale features of the geometry. In order to clarify that
the starting point is indeed in the universality class of the nonlinear sigma model on a narrow handle, in the
next subsection we will describe a deformation of the model which removes the extra $\IP^1$ in the geometry of
the $(2,2)$ model, and visibly smooths the connecting region between the $\phip$ and $\etap$ throats, but
preserves the RG trajectory of $\rho$.

\subsec{The (1,1) deformation}

Recall that the finite-size throat is incompatible with a winding
tachyon preserving $(2,2)$ supersymmetry. Armed with this
information, we seek out a deformation which preserves only
$(1,1)$ supersymmetry.  We will set it up so that the resulting
geometry of the low-energy vacuum manifold is determined by an
equation of exactly the form \desired\ discussed above.

Breaking $(2,2)$
to $(1,1)$ means that the chiral multiplets
each split into two real multiplets, and the
superpotential term is now a
full superspace integral of a real function
(which preserves $(2,2)$ if it's the
real part of a holomorphic function of
the formerly-chiral combinations).
This allows many more gauge invariant monomials
because $(1,1)$-preserving operators
can depend on both $\phi$ and $\bar \phi$.

We will modify the (2,2) model
of the previous subsection
 by adding the small 'real superpotential' term
$$ \delta w = -\mu ( \etatwo \bar \etatwo +
\phip \bar \etap + \bar \phip \etap  ).$$
By this we mean that we add to the lagrangian
\eqn\breaksNequalstwo{ L = q_+ q_- \left( \delta w \right) }
where $ q_\pm \equiv (1 / \sqrt{2}) ( Q_\pm + \bar Q_\pm )$ are
the two real supercharges we are going to preserve. For $(1,1)$
supersymmetry, the superpotential will include everything other
than the kinetic terms. Noting that
$$ \bar Q W = 0 \Longrightarrow
Q_+ Q_- W + \bar Q_+ \bar Q_- \bar W =
q_+ q_- ( W + \bar W), $$
the total real superpotential
(not including D-terms yet) is:
$$ w = m \left( \etatwo \phip \etap + {\overline{ \etatwo \phip \etap }}
\right) - \mu ( \etatwo \bar \etatwo + \phip \bar \etap + \bar
\phip \etap)  .$$ Note that $\mu$ must be real in order for the
action to be real. We are going to use the fact that the real
superpotential can still be differentiated with respect to complex
combinations of fields to figure out the F-term vacuum equations.
These equations are:
\eqn\Fetatwo{0= F_{\etatwo} = m \phip \etap - \mu \bar \etatwo }
\eqn\Fphi{0= F_{\phip } = m \etatwo \etap - \mu \bar \etap }
\eqn\Feta{ 0= F_{\etap } = m \etatwo \phip - \mu \bar \phip  }
and their complex conjugates. $F_{\phitwo} = 0$ trivially. Adding
\Fphi\ to \Feta\ determines $\etatwo$ in terms of the positively
charged fields (when $\phip, \etap \neq 0$):
\eqn\fixetatwo{ \etatwo = { \mu \over m} \left. { \overline{\phip
+ \etap} \over \phip +  \etap } \right. }
The second factor on the RHS is a phase, so this says that
$$ |\etatwo| = {\mu  \over m}.$$
Dividing \Fphi\ by \Feta\ gives
\eqn\relativephase{ {\phip \over \etap} = {\bar \phip \over \bar
\etap} }
which implies that this ratio is real.  Call it $x$:
\eqn\eliminateeta{ \etap = x \phip .}
Plugging these into \Fetatwo\ gives
\eqn\fixphi{ x |\phip|^2 = { \mu^2 \over m^2} ,}
which tells us that $x > 0$ by the reality of $\mu$ and $m$. Now
the D-term reads:
\eqn\Dtermeqn{ \left(x + {1 \over x}\right) { \mu^2 \over m^2} =
\rho + 2 {\mu^2 \over m^2}+ 2 |\phitwo|^2 }
Moving the $2 \mu^2/m^2 $ to the LHS gives
\eqn\victory{ \left( \sqrt x - {1 \over \sqrt x} \right)^2 {\mu^2
\over m^2} = \rho + 2 |\phitwo|^2 .}
Note that the double-valuedness from taking this square root is
not harmful to us since $x$ is always positive on the vacuum
manifold.

To recap:  $\etatwo$ is eliminated by \fixetatwo.
The equation \eliminateeta\ determines $\etap$
in terms of $x, \phip$.  The equation \fixphi\
eliminates $|\phip|$ in terms of $x$.
Because $\phip$ doesn't vanish on this branch of the
moduli space we can use the $U(1)$ gauge symmetry to
fix its phase.
The remaining variables are $x \in \IR_+$ and $\phitwo \in \IC$
related by the eqn \victory.
Solving this equation for the norm of $\phitwo$
gives a circle (the phase of $\phitwo$)
fibered over the $x$ direction.  Its radius is
$$
2 |\phitwo|^2 =
( \sqrt x - {1 \over \sqrt x} )^2 { \mu^2\over m^2} - \rho .
$$
This is the desired equation
\desired\ with $ w = \sqrt x - {1 \over \sqrt x} $.

In the IR, $\rho \to + \infty$\foot{
In the next subsection, we analyze the RG behavior
of the (1,1) theory and show
that nonzero $\mu$ does not
change the running of $\rho$ from the (2,2) result.},
and this equation
has no solutions for an interval of values of $x$ where
the RHS dips below zero.  There are thus two caps,
as there were for $\mu =0$.

In the UV, $\rho \to - \infty$ and the RHS is positive definite.
The radius of the $\phitwo$ circle is big
at the two ends ($ x \to \infty$ and $x \to 0$),
and has a minimum when $ x = 1$.

\subsec{Extra Vacua}

In the above analysis we have described the part of the vacuum
manifold that for large $|\rho|$ describes a geometrical target
space.  In general, there are other discrete massive vacua of the
system.  Physically, in the spacetime theory, these describe
subcritical-dimension target space components, which we will refer
to as `dust vacua' for want of a better name. From the point of
view of the world-sheet theory, the Witten index contributions
contained in the 10-dimensional geometrical regions jump from 0 in
the hyperboloid to -2 on the pair of caps.  The residual
contribution must be contained in the subcritical dust.

In general, we could start with some extra subcritical dust vacua
as well.  Indeed in the above model there is another solution of
the F-term equations when $\phip = \etap = \etatwo = 0 $. The
D-term equation becomes
\eqn\evilDterm{ 0 = \rho + 2 |\phitwo|^2 }
which in the UV, $\rho < 0$, has solutions. The gauge symmetry
fixes the phase of $\phitwo$, but there is a residual $\IZ_2 $
gauge symmetry, which acts on nothing.

In general there can also be vacua in which the scalar $\sigma$ in
the (2,2) gauge multiplet has a vev \WittenYC. We will discuss
these below, with the results that there are no such vacua for
$\rho < 0$, but for the IR limit $\rho
>0$ there are indeed $\sigma$ vacua.
Altogether, the Witten index for the sum of all components is conserved under the flow.

\subsec{GSO action on the extra `dust' vacua}

The question we want to answer in this subsection is the
following.  We start with a type II string theory, with its chiral
GSO projection.  The vertex operator \TsupvertII\ that we add is
of course invariant under this symmetry, and also does not break
the $Z_2\times Z_2$ global `quantum symmetry' corresponding to
this GSO projection (i.e. it is an NS-NS state).  It is therefore
interesting to ask how the GSO projection acts on the `dust'
components of the target space we obtain after the tachyon
condensation.

In particular, the isolated vacua (hereafter called 'dust') lead a priori to disconnected eight-dimensional
universes, which are flat at very weak coupling, and the obvious type II GSO projection in eight dimensions is
not modular invariant, while the type 0 one is. We will see how the discrete R symmetries in our linear model
act in a way which precisely reduces the type II GSO we start with to the type 0 GSO after the flow.  More
precisely, we obtain the eight dimensional target space in the dust components as an intermediate step available
in the RG flow; these subcritical theories are themselves tachyonic (they have relevant operators) and will
ultimately flow further down to 2 or fewer dimensions.

This works out as follows. In the linear model realization, the
dust vacua reside on the $\sigma$ branch of the field space, where
as mentioned above $\sigma$ is the scalar in the (2,2) vector
multiplet. (We can set $\mu = 0$ for this discussion, because it
is not important for the $\rho \to + \infty$ physics.) They arise
\WittenYC\ by solving the SUSY-preservation equations arising from
the quantum-corrected twisted chiral superpotential term
\eqn\twistedsp{ \int d\theta _+ d\bar \theta_- ~\tilde W = \int
d\theta_+ d\bar \theta_- ~ \left( t \Sigma + Q_{{\rm T}} \Sigma
\ln \Sigma \right)} where $ t = \rho + i \theta$, $ Q_{{\rm T}} $
is the net charge of the chiral fields, and the second term arises
from the chiral multiplets running in a loop.
Given that the chiral R-symmetries act by
\eqn\chiralR{ \theta_+ \mapsto e^{i \alpha_+} \theta_+, ~~~
\theta_-  \mapsto e^{i \alpha_-} \theta_- ,}
the presence of the first term implies that
$ \sigma $ transforms like
$$ \sigma \mapsto e^{ i \alpha_+ - i\alpha_-} \sigma .$$
The second term in \twistedsp\ reproduces the R-charge anomaly. Although the $U(1)_{{\rm axial}}$ is anomalous,
a $\IZ_2$ subgroup is preserved because the anomaly is even; this is important because it is by this $\IZ_2$
generator $g$ that the chiral GSO acts on this system.  That is, we have two independent $Z_2$ symmetries which
provide the chiral $\IZ_2\times\IZ_2$ symmetry by which we can orbifold to enforce the GSO projection
appropriate to the type II theory.  (By comparison, the type 0 GSO projection is obtained by the vectorlike
combination of these $\IZ_2$ symmetries.)

Now the key point is just that the sigma vacua are at values
of $\sigma$
such that
$$ 0 = {\del \tilde W \over \del \sigma } $$
which are
$$ \sigma_\pm = \pm e^{ |t|/2 }. $$
These two values are permuted by the action of $g$. Therefore, a
fundamental domain for the action of $g$ is obtained by forgetting
about one of the sigma vacua; the remaining theory is one isolated
vacuum, modulo the action of the vectorlike GSO projection, namely
one copy of 8-dimensional type 0.

\subsec{Renormalization properties of the (1,1) GLSM}

Since the deformation \breaksNequalstwo\ breaks
$N=(2,2)$ supersymmetry,
the usual nonrenormalization theorems
used in \WittenYC\
need not apply.
In order to ascertain whether
our linear model has the desired
flow, we analyze the running of
the $N=(1,1)$ couplings,
and of the possible dangerous terms that are now
allowed by symmetries.
Such dangerous terms include:
\eqn\dangerousterms{
q_+ q_- \left( \tilde m \bar \phitwo \phitwo +
 m_1 \left( \bar \phip \phip + \bar \etap \etap \right) \right).}
The generation of such terms at scales comparable to $e$ or $m$ or $\mu$ would destroy the vacuum manifold.

This analysis is facilitated by the
following observation.
Since $\phitwo$ does not appear in
any of the superpotential terms ($N=2$ or otherwise),
if we ignore the coupling to the vectormultiplet
there is a shift symmetry
\eqn\shiftsym{ \phitwo \mapsto \phitwo + \alpha }
with $\alpha$ an arbitrary constant superfield. This means that
potential terms that involve $\phitwo$ must be accompanied by a
positive number of powers of $e$ (an even number, by charge
conjugation symmetry) as well as $\mu$. Because of this, it is
actually quite difficult to generate the troublesome mass terms
\dangerousterms, particularly when combined with the
superrenormalizible nature of this field theory.

\itemized \ifig\dangerous{ a) a $\phi^4$ one-loop bubble; b) a
fermion loop; c) a $ \phi^2 \psi^2 $ bubble.
}{\epsfxsize3.0in\epsfbox{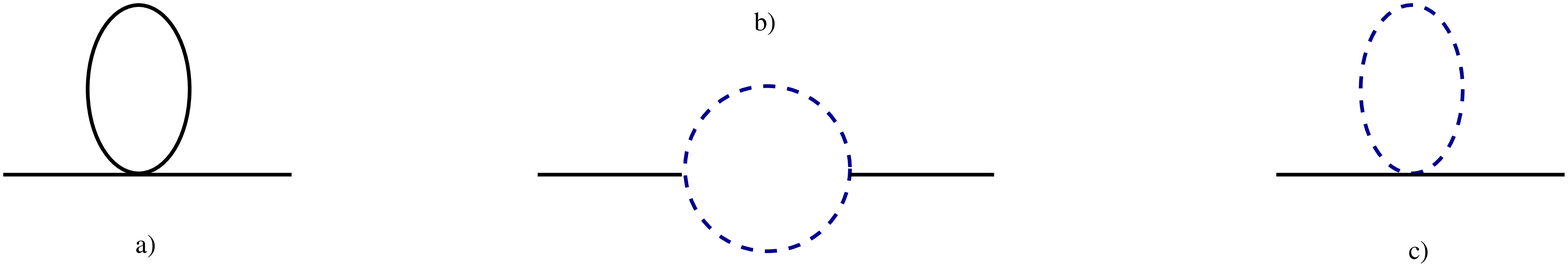}}

\itemaut{First of all, we find that in the theory of interest, the
only three classes of diagrams with two external scalars which can
have UV divergent parts are shown in \dangerous. This is just the
statement of superrenormalizibility of the theory, in detail for
this particular term of interest.}

\itemaut{There are no $|\phi|^2 |\psi|^2 $ couplings
in this model, so we never have to worry about a
diagram of type c.}

\ifig\manyloops{ Here are the two least-UV-finite diagrams that
renormalize the mass of $\phitwo$. The scalar propagators have an
orientation because the fields are complex and charged, which
isn't indicated. For this figure, $\phi$ means $\phip$. The two
vertices that have been used here are $ \mu \phip^2 \etatwo $ from
the '$\mu$-term' and its conjugate, and $ e^2 |\phitwo|^2
|\phip|^2 $ from the D-term. There are other diagrams where
$\phip$ is replaced by $\etap$ and there is another diagram of the
form of the first where the $\phip$ that hits the $\phitwo$ is
replaced by $\etatwo$ and $\etatwo$ and the other $\phip$ are both
replaced by either $\phip$ or $\etap$.
}{\epsfxsize4.0in\epsfbox{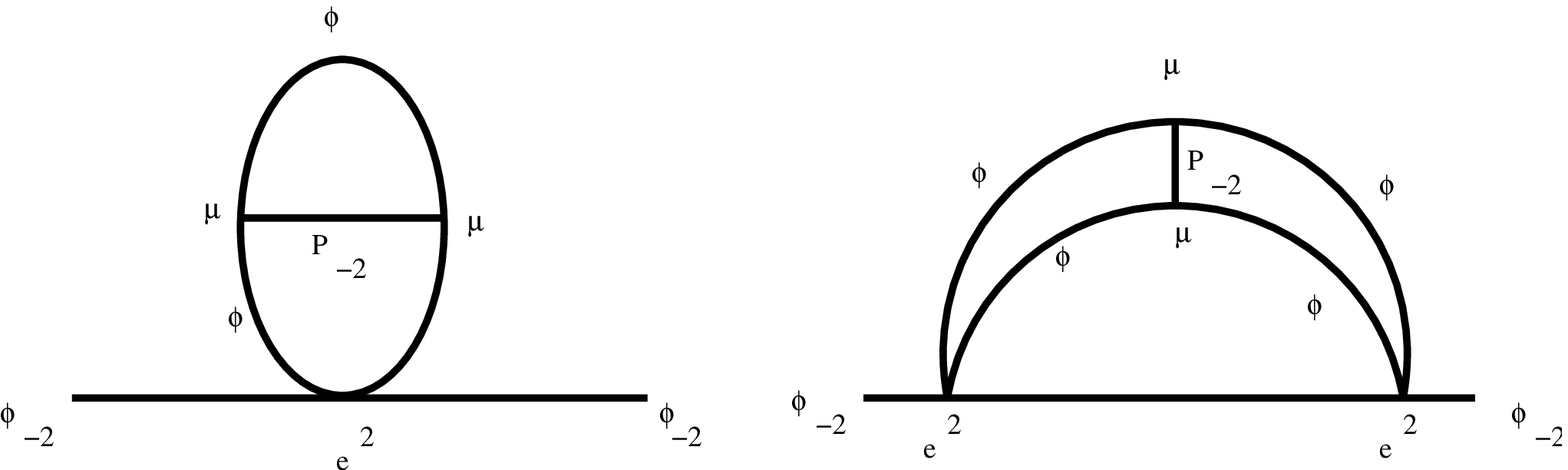}}

\itemaut{For the $\phitwo$ mass renormalization diagrams, any
putative extra divergence\foot{ By 'extra' we mean in addition to
the divergences arising from the renormalization of $\rho$ or of
the kinetic terms, which {\it do} appear here if you use the
D-term vertices to let the other scalars run in the loop in
diagram a -- but this is $\mu$-independent and therefore not
germane.
}
must be accompanied by
two powers of $e$
and
one power of $\mu$ (because $\mu$ is the only thing that breaks $N=2$).
There is no diagram of the form a or b that has this property.
The least non-divergent diagram that we've found with this property
has two loops, and four scalar propagators,
and is show in \manyloops.
}

\ifig\phimass{
The leading correction to the mass of
$\phip$ comes from this finite amplitude.
The vertices are from the
$ \mu \phip^2 \etatwo $ term and its conjugate.
}{\epsfxsize2.0in\epsfbox{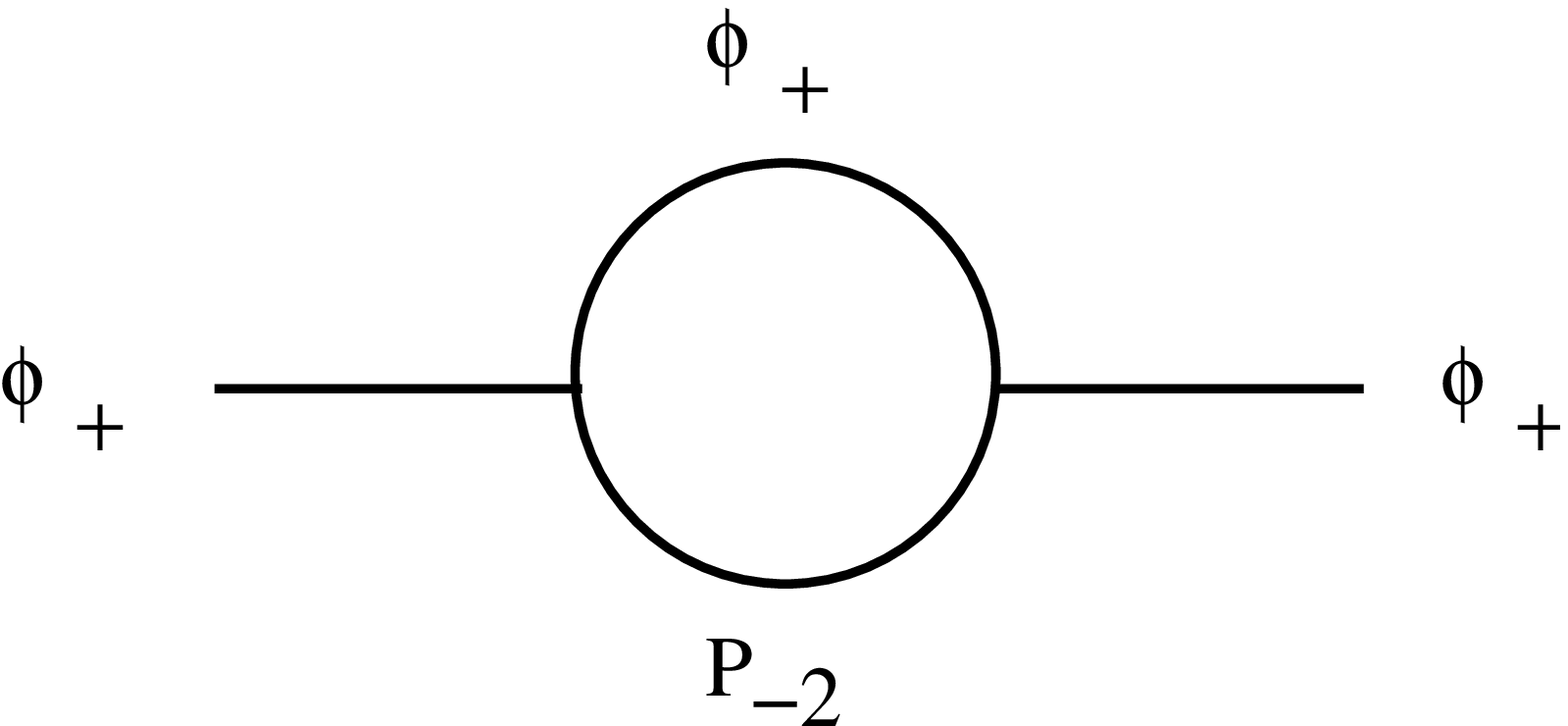}}
\itemaut{For the $\phip$ and $\etap$ mass renormalization,
and for the renormalization of $\mu$ itself,
we only know that it must depend on $\mu$.
There is still no $\mu$-dependent diagram of the form of diagram a.
For diagram b, the only option is to use the cubic vertices from the
$N=2$ superpotential to make $\phip$ turn into two fermions.
The simplest diagram of this form does indeed diverge
logarithmically in the UV,
but it doesn't depend on $\mu$.  This is just the wavefunction
renormalization of $\phip$ from loops of fermions
(it vanishes at zero external momentum,
and so doesn't renormalize the mass
-- an important fact for the $N=2$ LSM);
it is an innocuous correction to the kahler potential.
We can try to insert $\mu$-dependence by
using the 'mass vertices' from the $N=1$ superpotential
(note that there are {\it no} other terms involving
fermions that come from the $(N=2)$-breaking terms):
$$ L_f = \mu ( \psi^{\etatwo}_+ \bar \psi^{\etatwo}_-
+ \psi^{\etap}_+ \bar \psi^{\phip}_-
+ \bar \psi^{\etap}_+ \psi^{\phip}_-  )
+ \left({ + \leftrightarrow - }\right) .$$
But the powers of $\mu$ that you need to
insert are accompanied by as many extra propagators,
even one of which makes the integral UV finite.
Alternatively, we could just renormalize the fermion propagators
by summing the geometric series, and use these in diagram b.
The result is again that the $\mu$-dependent part
will not diverge.
The leading finite contribution to the $\phip, \etap$ masses
is shown in \phimass.}

\itemaut{Similarly, it is simple to show that
$\mu$ itself does not receieve any
UV divergent corrections.  Further,
there is no $\mu$-dependent
UV divergent correction to the FI parameter $\rho$.}

From the results 1-4, we can conclude that the beta function for
the dangerous couplings, which we collectively denote as $\tilde
m$, is of the form
$$ \beta_{\tilde m^2} = \sum_g {\del \over \del g} \delta \tilde m^2
\beta_g  , $$
where $\delta \tilde m^2 $ is the sum of the finite mass-shifts
discussed above. This is because there are no extra divergent
counterterms required to cancel these diagrams -- all of the
scale-dependence comes from the fact that these diagrams, and
hence the counterterms which cancel them, depend on the other
renormalized couplings. We expect that these finite diagrams have
the property that their dependence on the running coupling ${ \del
\delta \tilde m^2 \over \del \rho} $ is small when $|\rho|$ is
large. This is a consequence of the fact that the running coupling
$\rho$ appears, when at all, as a mass term, and that IR
enhancements are at worst logarithmic.

The final result of this analysis
is that it is possible to
fend off the dangerous terms \dangerousterms\
in the two large $|\rho|$ phases
by the addition of scale-independent counterterms.
The RG behavior of the
$(1,1)$ linear model
is therefore as described in
subsection A.2.

\subsec{A puzzle about the flow and its resolution}

At large negative $\rho$, the vacuum manifold seems to be a
one-sheeted hyperboloid of size $\rho$, independent of $\mu$. If
this vacuum manifold were in fact the target of the NLSM in the
IR, the following puzzle would arise. The string wound around the
waist of the hyperboloid would have a mass of order $l_s \sqrt
\rho \gg l_s$ and should therefore be very irrelevant. On the
other hand, $\rho$, which (as we have shown in \S A.1) turns on
this operator, has a constant, relevant beta function
$$ \beta_\rho = Q_{{\rm T}} = -2 .$$

The resolution of this puzzle involves two important subtleties of
the model. First of all, near the center of the throat, the
$\etatwo$ field is light. This is clear since before $\mu$ was
turned on, $\etatwo$ actually became massless at the point
($\phi_+ = \eta_+ =0$) where the extra $\IP^1$ was attached. This
is an indication of the presence of string-scale features in that
region.

More precisely, this opens up a fascinating possible loophole: for
small $\mu$, it does not cost much energy to move the $\etatwo$
field around. By moving it around, we will see that the energy of
the winding mode can be decreased from its apparent value if the
geometry really has small curvatures (of order ${1 \over |\rho|
}$) everywhere.

So we consider the energy of the wound string as a probe of how
big the cycle is. We pick a gauge where $\phitwo$ is the field
that winds
$$ \phitwo(\sigma) = e^{ i \sigma } \phitwo^0 ,$$
where $\sigma$ is the world-sheet space coordinate. The kinetic
energy density is something like $ e^2 | \phitwo^0 |^2 $. We add
this to the bosonic potential, with $\phip, \etap$ evaluated on
their vacuum solutions, $ \phip = x \mu, \etap = \mu / x $.
Further, we know that the wound string wants to be at the
narrowest part of the neck, $x=1$.

This gives the following answer for the energy, in string units:
$$
E( \etatwo^0, \phitwo^0 )
= e^2
\left(| \phitwo^0 |^2 + ( 2 {\mu^2 \over m^2} - \rho - 2 |\phitwo^0|^2
- 2 | \etatwo^0|^2 )^2 \right)
+ \mu^2 | {\mu\over m} - \etatwo^0 |^2 .
$$
Setting $\etatwo^0 = { \mu \over m}$,
its vacuum value, this gives
$$E(\etatwo = \mu/m) = e^2 \rho. $$
But taking into account the fact that
$\etatwo$ is light (\ie\ assuming $\mu^2 \ll e^2 $)
this actually has a much lower minimum, at
$$E = \mu^2 \rho. $$
Thus for small $\mu$ (as the (2,2) model becomes a better approximation), the winding mode may become tachyonic,
as required by the direction of flow of $\rho$ combined with its role multiplying the winding operator in the
action.

A related point is that from the GLSM instanton expansion, the tachyon vev (the coupling of the tachyonic
winding operator in the 2d action) is actually
$$ T(\rho) = f(\rho, \mu) e^{ \rho } . $$
Here $f(\rho)$ is a ratio of fermionic to bosonic one-loop determinants in the vortex background.  These
precisely cancel in the ${\cal N}=2$ limit, $f_{{\cal N}=2}=1$. This tells us that the beta function for $T$ is
actually
$$
\beta_T = \beta_\rho ~{ \del T \over \del \rho}
= Q_{{\rm T}} ~ ( { f' \over f} + 1 ) e^{ - \rho }.
$$
This can change sign as $\mu$ and $\rho$ vary.  Hence, for {\it large} $\mu$, far from the regime where ${\cal
N}=2$ results apply to a good approximation, the direction of flow may turn around, consistently with the
irrelevance of the winding operator for the large radius connected hyperboloid.

To summarize, the rewards of the linear sigma model analysis are the following:
\itemized
\itemaut{ The linear sigma model gives a global picture of the RG
flow, including the B-cycle direction, which corroborates the
analysis of the previous sections. }
\itemaut{ We found a {\it gauged} linear sigma model \WittenYC\
which `linearizes' the tachyon, in the sense that the winding
vertex operator becomes directly related to the vortices of the
GLSM. In such models, the condensation of vortices is controlled
by the Fayet-Iliopoulos parameter, whose renormalization
properties are well-understood. }
\itemaut{ The GLSM gives independent evidence for the existence of additional vacua, as in \refs{\othervacuarefs,\chicago}.}
\itemaut{The GLSM provides a nice picture of the mechanism by
which these lower-dimensional vacua manage to have a consistent
GSO projection.}


\vfill\eject
\appendix{B}{Towards a transformative hermeneutics 
of off-shell string theory}

\vskip0.7in

{\bf The Second Coming} -- W. B. Yeats

\newtoks\thisverbatim
\newtoks\everyverbatim
%
\let\preverbatim\medskip

\newcount\vrblin
\def\numvrb{\vrblin0
   \everypar{\advance\vrblin1
    \llap{\sevenrm\the\vrblin\quad}}}

\def\makeescape#1{\catcode`#1=0 }
\def\makeactive#1{\catcode`#1=13 }
{\makeactive\<
 }
\def\verbatim{\preverbatim\begingroup
   \tt\setupverbatim
   \the\everyverbatim\relax
   \the\thisverbatim\relax
   \verbatimgobble}
\def\setupverbatim{\makeactive\`%
   \let\*=*\makeescape\*
   \def\par{\leavevmode\endgraf}
   \obeylines\uncatcodespecials
   \obeyspaces}
{\obeyspaces\global\let =\ 
 \obeylines\gdef\verbatimgobble#1^^M{}%
 \makeactive\` \gdef`{\relax\lq}}
\def\uncatcodespecials{\def\do##1{%
   \catcode`##1=12 }\dospecials}
\makeactive\|
\def|{\bgroup\tt\setupverbatim
      \the\everyverbatim\relax
      \the\thisverbatim\relax
      \def|{\egroup\thisverbatim{}}}
%
%
%

\def\marginsert#1{
\vskip-0.361in\hskip3.6in\rlap{\footnotefont #1} 
\verbatim}


\verbatim
Turning and turning in the widening gyre
*endverbatim
\marginsert{
Vortex-induced}
The falcon cannot hear the falconer;
*endverbatim
\marginsert{causal disconnection}
Things fall apart; the centre cannot hold;
*endverbatim
\marginsert{follows from tachyon condensation,}
Mere anarchy is loosed upon the world, 
*endverbatim
\marginsert{creating an explosion of}
The blood-dimmed tide is loosed, and everywhere
*endverbatim
\marginsert{decay products, and opening}
The ceremony of innocence is drowned;
*endverbatim
\marginsert{a can of worms.
}
The best lack all convictions, while the worst
Are full of passionate intensity.
*endverbatim
\marginsert{
}

Surely some revelation is at hand;
Surely the Second Coming is at hand.
The Second Coming! Hardly are those words out
When a vast image out of Spiritus Mundi
Troubles my sight: somewhere in sands of the desert
A shape with lion body and the head of a man,
A gaze blank and pitiless as the sun,
Is moving its slow thighs, while all about it
Reel shadows of the indignant desert birds.
The darkness drops again; but now I know
That twenty centuries of stony sleep
Were vexed to nightmare by a rocking cradle,
And what rough beast, its hour come round at last,
Slouches towards Bethlehem to be born?
*endverbatim

\listrefs
\end